\documentclass[]{article}	

\usepackage[backend=biber,style=numeric,sortcites=true,
sorting=none,
url=false,eprint=false]{biblatex}

\usepackage{csquotes}
\usepackage{fix-cm}	 
\usepackage{fullpage}

\usepackage{fancyhdr} 
\usepackage{lastpage} 

\usepackage{float}
\usepackage{graphicx}
\usepackage{amsmath,amssymb,amsfonts,amsthm,mathtools} 
\usepackage[english]{babel} 
\usepackage[colorlinks=true, linkcolor=blue, citecolor=blue, filecolor=blue,
runcolor=blue, urlcolor = blue]{hyperref}
\usepackage[svgnames, table]{xcolor} 

\usepackage{subcaption}

\graphicspath{ {./figures/} }

\newcommand{\mb}[1]{\mathbf{#1}}

\usepackage{lettrine} 
\newcommand{\initial}[1]{ 
\lettrine[lines=3,lhang=0.3,nindent=0em]{
\color{DarkGoldenrod}
{\textsf{#1}}}{}}

\definecolor{atz_table1}{rgb}{0.85, 0.85, 0.85}
\definecolor{atz_table2}{rgb}{0.8, 0.8, 0.8}
\definecolor{atz_table3}{rgb}{0.75, 0.75, 0.75}

\newcommand{\refp}[1]{(\ref{#1})}

\usepackage{titling} 

\newcommand{\HorRule}{\color{DarkGoldenrod} 
  \rule{\linewidth}{1pt}} 

\pretitle{\vspace{-80pt} \begin{flushleft} \HorRule \fontsize{12}{12} 
  \color{DarkRed} \selectfont}

\title{Coarse-Grained Methods for Heterogeneous Vesicles with Phase-Separated Domains: \\
Elastic Mechanics of Shape Fluctuations, Plate Compression, and Channel Insertion.}

\posttitle{\par\end{flushleft}}

\preauthor{\vspace{-5pt} \begin{flushleft}\fontsize{12}{12} 
\large  \color{DarkRed}} 
\author{David A. Rower$^{1}$, P. J. Atzberger$^{2,3}$ }

\postauthor{\fontsize{12}{12} \small \color{Black} \\

$[1]$ Department of Physics, University of 
California Santa Barbara (UCSB); \\
$[2]$ Department of Mathematics, University of 
California Santa Barbara (UCSB); \\ atzberg@gmail.com; http://atzberger.org/ \\
$[3]$ Department of Mechanical Engineering, University of 
California Santa Barbara (UCSB); \\
\vskip 1.5em
\lfoot{\footnotesize * Work supported by DOE Grant ASCR PhILMS DE-SC0019246 
and NSF Grant DMS-1616353.}
\end{flushleft} 
\vspace{-18pt} \HorRule}

\date{}

\begin{document}

\maketitle 

\vspace{-1cm}

\thispagestyle{fancy}

\initial{W}\textbf{{e develop coarse-grained particle approaches for
studying the elastic mechanics of vesicles with heterogeneous membranes having
phase-separated domains.  We
perform simulations both of passive shape fluctuations and of active systems
where vesicles are subjected to compression between two plates or subjected 
to insertion into narrow channels.  Analysis methods are developed for 
mapping particle configurations to continuum fields with 
spherical harmonics representations.  
Heterogeneous vesicles are found to exhibit rich
behaviors where the heterogeneity can amplify surface two-point correlations,
reduce resistance during compression, and augment vesicle transport times
in channels.  The developed  methods provide general approaches for
characterizing the mechanics of coarse-grained heterogeneous systems taking
into account the roles of thermal fluctuations, geometry, and phase
separation.}}

\setlength{\parindent}{5ex}

\section{Introduction \label{sec:introduction}} 
{Biological membranes consist of complex heterogeneous mixtures of
different lipid species, proteins, and other small molecules~\cite{Singer1972,
Lingwood2010, Alberts2002}.  Membrane heterogeneity is important to organize
cellular structures and mechanical responses~\cite{AriolaFluidity2009,
KellerPhaseSep2003, SemrauMemMemInteractions2009, GrovesBendingPhaseSep2010,
BassereauShape2014}.  This includes the insertion, assembly, and activation of
membrane-proteins such as channels, receptors, or anchoring sites of the
cytoskeleton~\cite{Alberts2002, LeeMembraneProtein2003, EdidinRaftsReview2003,
LeeMembraneProteinInteractions1977, DohertyMembraneCytoskeleton2008}.  Phase
domains also play a role in the initiation of the formation of buds and
endocytosis~\cite{Kirchhausen2000,SackmannBudding1993, Baumgart2003,
McMahon2005,osterBudding1989}, and in the local control of
diffusivity~\cite{MarrinkDiffusionRafts2010, Groves2018,
Korlach2005,BassereauShape2014}, fluidity~\cite{AriolaFluidity2009}, or bending
moduli~\cite{Baumgart2003,Usery2018,GrovesBendingPhaseSep2010}.  For
applications and as model physical systems, synthetic soft membranes also have
been introduced consisting of self-assembled particles with phases that can
form sheets and other structures
~\cite{DogicColloidalMembrane2010,Dinsmore2002,Bollhorst2017,DogicColloidalMembraneRafts2014}.
} 

{Heterogeneous membranes can exhibit rich behaviors and geometries arising
from the phase-separated
domains~\cite{Baumgart2003,KellerPhaseSep2003,KimSangwoo2015Hvaa}.  For
spherical vesicles, phase separation kinetics and surface transport have been
studied experimentally
in~\cite{Baumgart2003,KellerPhaseSep2003,KimZasadzinskiSquires2011}.  To better
understand the mechanics of heterogeneous membranes, we develop coarse-grained
simulation approaches and analysis methods.    We consider the case of vesicles
with phase-separated domains having different preferred curvatures.
Simulations are performed to study the impact of heterogeneity both on passive
shape fluctuations and on vesicles subjected to active deformations. To
characterize mechanical responses, analysis methods are developed for mapping
particle configurations to continuum surface representations. Lebedev sampling
and spherical harmonic expansions are used to develop spectral analysis methods
for studying passive shape fluctuations and other geometric changes.}

{To investigate active deformations, simulation and analysis methods are
developed for vesicles subjected to compression between two plates and
subjected to insertion into narrow channels.  We find heterogeneous vesicles
can exhibit mechanical responses differing significantly from the homogeneous
case.  For heterogeneous vesicles, the phase domains are found to amplify
surface two-point correlations.  For vesicle compression, it is found that the
phase-domains can rearrange to accommodate curvature in membrane bending to
reduce energetic costs.  In channel insertion and
transport,  the heterogeneity is found to have mixed effects depending on the
circumstances which can either decrease or increase insertion and transport
times.  }

\lfoot{}

There have been many continuum mechanics and coarse-grained methods
developed for investigating multi-component membranes.  {At the continuum
mechanics level, these include~\cite{Lipowsky2003,Juilcher1993,Andelman1992,
LaradjiGinzburgLandauModel1991,Kumar1998,Boal1992,Kamal2012}.  At the
coarse-grained level, there are models of different levels of detail.  Models
incorporating some structure of the individual lipids
include~\cite{Marrink2007,Wang2010,Revalee2008,Goetz1999,Goetz1998a,
Imparato2004} and implicit-solvent models~\cite{Farago2003,Cooke2005a,Cooke2005,
Brannigan2006,Noguchi2001,Wang2010}. Models at a more coarse-grained level
employing a single-bead to describe a cluster of lipids
include~\cite{Laradji2004,Yuan2010,LaradjiBudding2005,Drouffe1991,Karniadakis2014}.
Related models and methods for use in dynamic studies have been developed
in~\cite{AtzbergerBilayerVesicle2013,LaradjiDPD2009,Karniadakis2014,
AtzbergerGrossMeshless2020,AtzbergerRower2022}. 
In general, coarser models have the advantage of computationally facilitating access
to larger length and time scales, but with the trade-off with the level 
of physical resolution at small scales~\cite{Muller2006,LaradjiReview2011}.}

As a basis for our computational studies, we use the single-bead
implicit-solvent coarse-grained model of Yuan \textit{et al.}~\cite{Yuan2010}.
In this approach, the membrane is modeled by a collection of orientable beads
having an additional director degree of freedom.  For biological membranes, the
beads can be thought of as representing small patches of lipids and their
orientation as taking into account on a coarse-grained level the collective
molecular-level orientation order within the patch.  For synthetic colloidal
membranes, the director-bead model can be interpreted as capturing the
effective physical interactions of distinct colloidal particles that are
orientable and with the sterics handled by spherical repulsive interactions.
{In our work, we use the molecular dynamics simulation framework referred
to as Large-scale Atomic/Molecular Massively Parallel Simulator 
(LAMMPS)~\cite{PlimptonLAMMPS1995}.  Related to~\cite{YuanLAMMPS2017} and
our prior work~\cite{AtzbergerLAMMPS2016}, we develop stochastic
time-step integration methods.}

{We also develop and implement custom analysis tools mapping particle
configurations to continuum fields for characterizing behaviors of the models.
The particle-based methods, combined with our analysis tools, yield approaches
that can avoid some of the cumbersome aspects of continuum formulations, which
often require formulations drawing on differential geometry and development of
numerical discretizations~\cite{AtzbergerGrossHydro2018,
AtzbergerGrossMeshless2020,AtzbergerRower2022}.  In this way, we are able to handle similar 
geometric contributions and physical phenomena for the phase-separation and
mechanics.  Much of the modeling is then deferred to the particle-level 
resolution and parametrization.  Our analysis tools provide general 
methods for relating such particle-based approaches to continuum-level concepts.  
}

{To quantitatively characterize our coarse-grained models and simulation
results, we develop spectral analysis methods based on the continuum mappings.
The vesicle shape is characterized using spherical harmonics expansions
of the mapping operator constructed using Lebedev quadratures~\cite{Lebedev1976,
Lebedev1999, AtzbergerSoftMatter2016}.  Results from statistical mechanics 
are used to analyze the passive shape fluctuations to estimate elastic
bending moduli and other properties.  This is studied when varying 
the vesicle size and phase concentrations.}

{In Section~\ref{sec:CG_model}, the coarse-grained model and numerical
methods are discussed.  In Section~\ref{sec:methods}, the simulation approaches
are discussed.  In Section~\ref{sec:spectral_analysis}, the spectral analysis
methods are discussed.  In Section~\ref{sec:results}, the results are discussed
for the simulations of active deformations when vesicles are subjected to
compressed between two plates or subjected to insertion into channels.  The
results show a few mechanisms by which phase-separated domains can impact the
mechanical responses of heterogeneous vesicles relative to the homogeneous
case.  The developed approaches also can be used to study  
other experimental measurements and simulations for phenomena within
heterogeneous membranes.  
}

\section{{Coarse-Grained Approach}}
\label{sec:CG_model}
\subsection{{Single-Bead Coarse-Grained Model for Heterogeneous Membranes}}
We investigate the
mechanics of heterogeneous vesicles at the level of coarse-grained models and
continuum mechanics.  As a mesoscopic description of the phase separation and
mechanics, we utilize the single-bead coarse-grained model of Yuan \textit{et
al.}~\cite{Yuan2010}.  In this approach, the membrane is modeled by a collection
of orientable spherical beads that have both a translational degree of freedom
$\mb{r}_i$ for the center-of-mass location and a rotational degree of freedom
$\mb{n}_i$ for the direction of orientation.

\begin{figure}[H] \centering
\includegraphics[width=0.99\columnwidth]{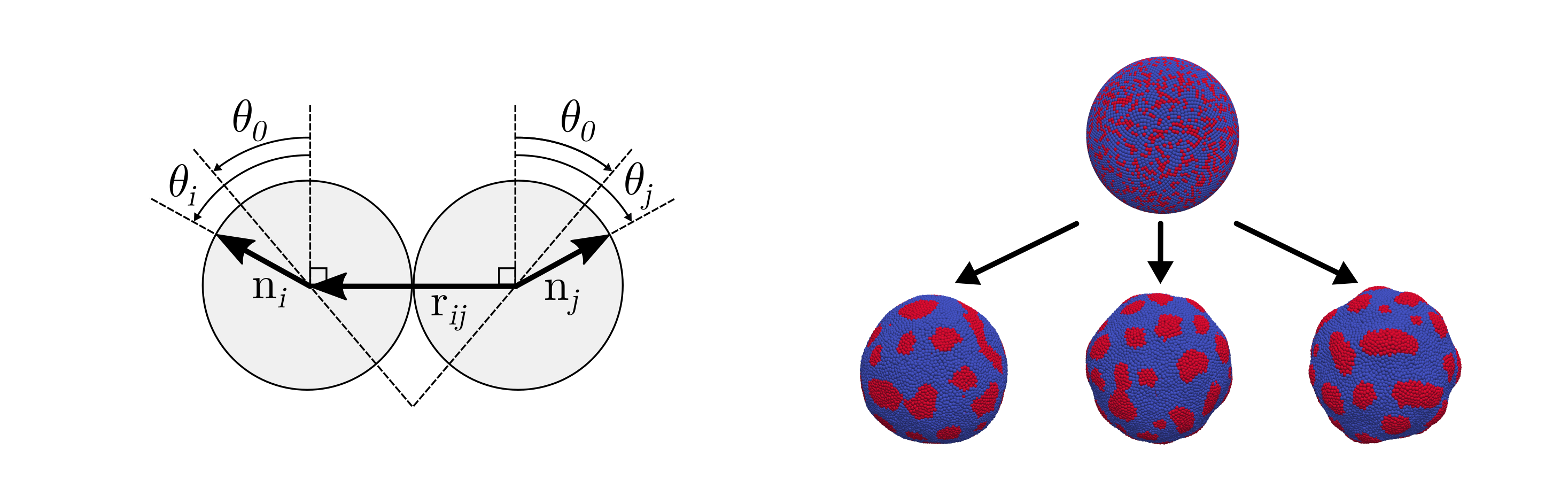}
\caption{{Coarse-grained particles and their interactions.  The 
particles are at locations 
$\mb{r}_i$ and $\mb{r}_j$. The interactions depend on both the relative
displacement $\mb{r}_{ij}$ and the orientations $\mb{n}_i$, $\mb{n}_j$.  The
$\theta_i$, $\theta_j$ give the in-plane angles.  The interaction energy 
is given in equation~\refp{Potential}, with minimization occurring when 
$|\mb{r}_{ij}| = \left(2\right)^{1/6}\sigma$ and 
$\theta_i = \theta_j = \theta_0$ \textit{(left)}.
These interactions can have different affinities and preferred curvatures
to drive phase-separation and shape changes in heterogeneous vesicles 
\textit{(right)}} } 
\label{fig:single_bead_model}
\end{figure}

For lipid membranes, the beads can be thought of representing at the
coarse-grained level a small patch of lipids with the orientation $\mb{n}_i$
serving as the local order parameter for the collective alignment of lipid
molecules within the patch.  This allows at the coarse-grained level to account
for molecular interactions arising from hydrophilic-hydrophobic effects driving
membrane assembly~\cite{Israelachvili2011}.
One could also interpret the model as a representation for a colloidal membrane
with the beads representing individual colloidal particles having
orientation-dependent interactions and spherical sterics.  In either case, the
orientation degrees of freedom $\mb{n}_i$ provide an important way to capture
broken symmetries in the physical system that drive assembly of two-dimensional
membrane sheets.

{The parameters and quantities are referenced to the non-dimensional
Lennard-Jones (LJ) characteristic scales $(\epsilon,\sigma,m)$.  
These are energy $\epsilon$, length $\sigma$, and mass $m$.  
We use characteristic time-scale $\tau
\sim \sqrt{m\sigma^2/\epsilon}$.  As in dimensional analysis, 
in this way our
simulations reflect behaviors of all realizable physical systems in the
equivalence class that are dynamically symmetric with respect to the
characteristic LJ references.  For instance, one can convert our results for the
coarse-grained system to particular physical units by taking $\sigma$ to be the
length-scale of a cluster of lipids, $m$ to be the mass of a cluster of lipids,
and $\epsilon$ based on the temperature (thermal energy) of the system.}
{To account for the interactions between two beads at location $\mb{r}_i$ and
$\mb{r}_j$, unit vectors are introduced to account for the orientation
$\mathbf{n}_i$ and $\mathbf{n}_j$.  The relative displacement between
the two beads is denoted by $\mathbf{r}_{ij} = \mathbf{r}_i - \mathbf{r}_j$. In these
interactions, the beads have a preferred angle of alignment, which we denote by
$\theta_0$.  The interaction energy is minimized when 
$\theta_i = \theta_j = \theta_0$.
The interactions between beads is illustrated in Figure~\ref{fig:single_bead_model}.}

{To account for interactions between pairs of beads, a potential energy $U$ 
is used that depends on both the relative positions and orientations.  This has the
general form}
\begin{equation}
U(\mathbf{r}_{ij},\mathbf{n}_{i},\mathbf{n}_{j}) =
\begin{dcases}
    u_{R}(r)+[1-\phi(\mb{s})]\epsilon,\quad r<r_{b} \\
    u_{A}(r)\phi(\mb{s}), \quad r_{b}<r<r_{c},
\end{dcases}
\label{Potential}
\end{equation}
{where $\mb{s} =
(\mathbf{\hat{r}}_{ij},\mathbf{n}_{i},\mathbf{n}_{j})$.  The $u_R$ is a repulsive potential
and $u_A$ is an attractive potential with $\phi$ mediating the transition between these
behaviors.  The $\phi$ is taken to depend on the relative orientations of the beads
as}
\begin{eqnarray}
\phi(\mathbf{\hat{r}}_{ij},\mathbf{n}_{i},\mathbf{n}_{j}) =
  1+\mu[a(\mathbf{\hat{r}}_{ij},\mathbf{n}_{i},\mathbf{n}_{j})-1],
\label{phi}
\end{eqnarray}
where
\begin{equation}
a = (\mathbf{n}_{i}\times\mathbf{\hat{r}}_{ij})\cdot
  (\mathbf{n}_{j}\times\mathbf{\hat{r}}_{ij})
  -\sin\theta_{0}(\mathbf{n}_{j}-\mathbf{n}_{i})\cdot
  \mathbf{\hat{r}}_{ij}-\sin^{2}\theta_{0}.
\label{a}
\end{equation}
{The repulsive steric
interactions $u_{R}$ are modeled by a 4-2 Lennard-Jones (LJ) potential energy}
\begin{eqnarray}
u_{R}(r)=\epsilon\bigg[\Big(\frac{r_{b}}{r}\Big)^{4} -
  2\Big(\frac{r_{b}}{r}\Big)^2\bigg], \hspace{0.5cm} r \leq r_{b},
\label{rPotentialSmall}
\end{eqnarray}
with $u_R(r) = 0$ for $r > r_{b}$. {The value $r_{b} = \sqrt[6]{2}\sigma$ 
is used for the effective size of a bead.}  The attractive interactions $u_{A}$ are modeled by
the potential energy
\begin{eqnarray}
u_{A}(r)=-\epsilon\cos^{2\zeta}\bigg[\frac{\pi}{2}
  \frac{(r-r_{b})}{(r_{c}-r_{b})}\bigg], \hspace{0.5cm} r_{b}<r<r_{c}.
\label{rPotentialBig}
\end{eqnarray}
{The parameter $\zeta$ allows for tuning the range of the particle
attractive forces.  The $\epsilon$ controls the strength of the bead-bead
interactions and in general is chosen differently according to the different
bead interaction pairings.}  The interactions are truncated at the critical
cut-off length $r_c=2.6\sigma$.

{The $u_R$ potential models repulsion between the beads with a preferred
separation distance given by a weak long-range attraction similar to a
Lennard-Jones potential.  The $u_A$ potential models a stronger attraction
chosen to have a wide energy-well which allows for beads to exchange cohesively
within the sheet promoting fluid-like behaviors.  The $u_A$ potential 
captures the long-range interactions arising in hydrophobic-hydrophilic
effects~\cite{Israelachvili2011}.  Further discussions on how fluid-phases
arise in membranes, and the roles of such potentials in coarse-grained models,
can be found in the works by Farago~\cite{Farago2003} and Deserno et
al.~\cite{Cooke2005}. } {A summary of the key parameters of the model is
given in Table~\ref{table:model_param_meanings}. The default values used
throughout the simulation studies are given in
Table~\ref{table:general_parameterizations}.}

\begin{table}[H]
\centering
\begin{tabular}{|l|l|}
\hline
\rowcolor{atz_table1}
Parameter & Description \\ \hline
$r_{b}$    & Distance for LJ minimum.        \\
$r_c$      & Force cutoff distance.          \\
$\zeta$    & Repulsive strength exponent.    \\
{$\epsilon$} & {Interaction strength.} \\
$\theta_0$ & Preferred relative orientation. \\
$\mu$      & Strength of orientation penalty.\\
\hline

\end{tabular}
\caption{\label{table:model_param_meanings} {Parameters of the coarse-grained model.  The values for the pairwise interactions are given
in Table~\ref{table:general_parameterizations}.}}
\end{table}

\subsection{Stochastic Time-Step Integrator}
{The simulations of the membrane are performed in the canonical NVT ensemble,
where there is a constant number of particles $N$, fixed volume $V$,
and constant temperature $T$~\cite{Reichl1998,Frenkel2002}.  We use the 
Langevin dynamics} 
\begin{align}
m\frac{d\mb{v}}{dt} = \mathbf{F},\; \frac{d\mb{r}}{dt} = \mb{v},\; I\frac{d\mb{w}}{dt} = \mathbf{t},\; \frac{d \mb{q}}{dt} = \mb{w},
\end{align}
{where}
\begin{equation}
\label{equ:langevin_fix}
\mathcal{F} = [\mathbf{F},\mathbf{t}]  =
\mathbf{F}_{U} + \mathbf{F}_{F} + \mathbf{F}_{T}.
\end{equation}
{The $\mathbf{r}$ denotes the collective bead positions and $\mathbf{v}$
the collective bead velocities.  The $m$ is the mass and $I$ plays a role
similar to the moment of inertia. The $\mathbf{w}$ is the tangential velocity of the
directors, and $\mathbf{q}$ is the collection of quaternion~\cite{Judson1997}
vectors for the rotational description of the orientation of the beads.  The
$\mb{F}$ is the force.  The $\mathbf{t}$ are the forces associated with the
director degrees of freedom, similar to a torque.   While the director forces
will sometimes be referred to as torques for brevity, these quantities
technically would require additional cross-products.}

The $\mathbf{F}_{U} = -\nabla_{\mathbf{r},\mathbf{n}}U$ denotes force from the
potential $U$ in equation~\refp{Potential} with collective gradient with respect to
both the translational and rotational degrees of freedom.  Expressions
are given for computing forces and torques in Appendix~\ref{appendix:gradients}.  The
friction $\mathbf{F}_F$ is modeled by the forces $\mathbf{F}_F =
[\mathbf{F}_F^{transl},\mathbf{F}_F^{rot}]$.  {The $\mathbf{F}_F^{transl} = -
({M}/{\nu_r}) \mathbf{v}$ involving $M$ for the mass of a particle and $\nu_r$
for the decay timescale.   Similarly,  $\mathbf{F}_F^{rot} = -({I}/{\nu_n})
\mathbf{w}$ involving $I$ for the director mass moment and $\nu_n$ for the decay
time-scale.  The thermal fluctuations contribute through the stochastic force
$\mathbf{F}_{T} = [\mathbf{F}_{T}^{transl},\mathbf{F}_{T}^{rot}]$, which
represents interactions with the solvent and other implicit degrees of freedom
when the temperature is $T$.  The thermal forces have strength determined by
fluctuation-dissipation balance~\cite{Reichl1998}.  The translational thermal
forcing is $\mathbf{F}_{T}^{transl}(t) \sim \sqrt{k_B T M} \xi(t)$, and the
rotational thermal forcing is $\mathbf{F}_{T}^{rot}(t) \sim \sqrt{k_B T I}
\tilde{\xi}(t)$. Formally, the $\xi(t)$ and $\tilde{\xi}(t)$ are independent
white-noise Gaussian processes~\cite{Gardiner1985} with $\langle \xi(t)\xi(t')
\rangle = \delta(t - t')$ and $\langle \tilde{\xi}(t)\tilde{\xi}(t') \rangle =
\delta(t - t')$.  This gives the stochastic update of the system. 
}

{We develop these
Langevin thermostats and associated inertial stochastic time-step integrators
for both the translational and rotational degrees of freedom.  Our approach
shares similarities with Velocity-Verlet integrators~\cite{Frenkel2002}, in which 
half-time steps are used to update the velocity $\mb{v}$ and 
angular velocity $\boldsymbol{\omega}$.  These are then used to update \
the translational and orientational configuration degrees of freedom 
$\mb{r}, \mb{n}$.  While the presence of the friction and thermostatting forces 
no longer give dynamics with strict time-reversibility, using the Verlet-like 
integrator does still help with the contributions of the conservative terms and 
with numerical stability~\cite{Frenkel2002}.}

{Our stochastic time-step integrator has three main stages.  The first stage
updates over a half time step the momentum associated with the translational
degrees of freedom.  The second stage updates over the full time step the 
translational degrees of freedom.  The third stage 
updates the rotational degrees of freedom.  The forces and torques are recomputed
with the momentum of the translational
and rotational degrees of freedom updated over the remaining half time step.  
The friction and stochastic terms contribute during updates as part of the 
forces and effective torques.}
{The translational degrees of freedom are updated using}
\begin{eqnarray}
\begin{aligned}
    \mathbf{v}^{n+\frac{1}{2}} =
      \mathbf{v}^{n}+\frac{\Delta t}{2}\mathbf{a}^{n}, \;\;\;\;
    \mathbf{r}^{n+1} = \mathbf{r}^{n} + \Delta t \mathbf{v}^{n+\frac{1}{2}} 
\end{aligned}.
\label{rIntegration}
\end{eqnarray}
{The rotational degrees of freedom are updated using}
\begin{eqnarray}
\begin{aligned}
    \mathbf{w}^{n+\frac{1}{2}} =
    \mathbf{w}^{n}+\frac{\Delta t}{2}\mathbf{A}^{n}, \;\;\;\;
    \boldsymbol\omega^{n+\frac{1}{2}} \leftarrow \mathbf{w}^{n+\frac{1}{2}},
      \quad \mathbf{\hat{n}}^{n} \leftarrow \mathbf{q}^{n},\\
    \mathbf{\hat{n}}^{*,n+1}=\mathbf{\hat{n}}^{n} +
      \Delta t (\boldsymbol\omega^{n+\frac{1}{2}}\times \mathbf{\hat{n}}^{n}),\;\;\;\;
    \mathbf{\hat{n}}^{n+1}=
      \frac{\mathbf{\hat{n}}^{*,n+1}}{\lvert\mathbf{\hat{n}}^{*,n+1}\rvert}, \;\;\;\;
    \mathbf{q}^{n+1} \leftarrow \mathbf{\hat{n}}^{n+1}.
\end{aligned}
\label{nIntegration}
\end{eqnarray}
{The momentum is updated using}
\begin{eqnarray}
\begin{aligned}
\mathbf{v}^{n+1} = \mathbf{v}^{n+\frac{1}{2}}+\frac{\Delta t}{2}
\mathbf{a}^{n+1},\;\;\;\;
\mathbf{w}^{n+1} = \mathbf{w}^{n+\frac{1}{2}}+\frac{\Delta t}{2}
\mathbf{A}^{n+1}
\end{aligned}.
\label{finalIntegration}
\end{eqnarray}
{The $\mathbf{\hat{n}}$ denotes the collection
of orientation director vectors associated with the beads.   The 
$\mathbf{a}^{n}={\mathbf{F}^n}/{M}$ denotes the
acceleration of the beads, where $\mathbf{F}^n$ is the force 
from equation~\refp{equ:langevin_fix}.  The $M$ is the mass of
the beads. The angular acceleration is denoted by $\mathbf{A}^n = \mathbf{t^n}/I$,
where the $\mathbf{t}^n$ are the forces associated with the director degrees of
freedom, similar to a torque from equation~\refp{equ:langevin_fix}.  
The $I$ plays a role similar to the moment of
inertia.  The $\omega$ denotes the angular velocity tangent the
directors obtained from the rate of change of the quaternions 
$\mb{w}$.  Each of these updates use 
the forces and torques acting on the
system at time ${n+1/2}$
to obtain $\mathbf{a}^{n+1}$, $\mathbf{A}^{n+1}$.}

{The thermal fluctuations contribute through the stochastic force
$\mathbf{F}_{T} = [\mathbf{F}_{T}^{transl},\mathbf{F}_{T}^{rot}]$, which
represents interactions with the solvent and other implicit degrees of freedom
when the temperature is $T$.  The thermal forces have strength determined by
fluctuation-dissipation balance~\cite{Reichl1998}.   In the discretization 
in time, we use $\xi(t_n)
\approx \zeta_n$ where $\zeta_n$ are Gaussian variates with mean $0$ and
covariance $\langle \zeta_n\zeta_m \rangle = \delta_{mn}/\Delta{t}$, where
$\Delta{t}$ is the time-step and $\delta_{mn}$ the Kronecker $\delta$-function.
This contributes to the discrete stochastic updates of the system each time-step.  }

{To validate the models, comparisons were made with statistical mechanics 
for the Maxwellians
of the rotational and translational velocities~\cite{Reichl1998}. For
further validation, we observed the domain interface lengths follow the inverse
power law $L \sim t^{-\alpha}$, where we find $0.2 \leq \alpha \leq 0.3$. 
It was observed that stalling can occur for the domain coarsening in a 
manner positively
correlated with the interaction parameter $\theta_{hc,hc}$ defined in
Table~\ref{table:general_parameterizations}.  It was found the 
cross-species well-depth needed to be reduced in order to 
help drive phase-separation in our models.  Our results 
are consistent with previous simulation results
in~\cite{YuanMemMediated2011}.  We
implemented the single-bead models
by introducing custom stochastic time-step integrators 
and force interaction laws within the molecular dynamics software 
LAMMPS~\cite{PlimptonLAMMPS1995,AtzbergerLAMMPS2016}.}

\section{{Simulation Approach}} 
\label{sec:methods}

\subsection{Parametrization} To obtain stable vesicles exhibiting phase
separation in a fluid phase membrane, we performed exploratory simulation
studies over the parameters.  To model phase separation we introduced a weight
for the interactions between the high-curvature ($hc$) beads and the
lower-curvature base ($b$) phase. {The weight factor used was $0.65$.  To
characterize the relative concentration of the high-curvature phase, the
percentage $n_{hc}$ is reported for the beads representing the phase 
relative to the total number of beads.  For all of our simulations, 
default base-line parameters are given in 
Table \ref{table:general_parameterizations}.}
\begin{table}[H]
\centering
\begin{tabular}{|lllll|}
\rowcolor{atz_table1} \hline
Pair & $\theta_0$ & $\mu$ & $\zeta$ & $\epsilon$ \\ \hline
\textit{b},\textit{b}    & 0.0               & 6.0   & 4.0  & 1.0        \\
\textit{b},\textit{hc}   & $\theta_{hc, hc}$ & 3.0   & 4.0  & 0.65       \\
\textit{hc},\textit{hc}  & $\theta_{hc, hc}$ & 6.0   & 4.0  & 1.0        \\
\hline
\end{tabular}
\caption{\label{table:general_parameterizations}{Parameters for
the coarse-grained model.  The values are expressed in terms of the
Lennard-Jones (LJ) characteristic scales.  The $b$ refers to the base-phase
and $hc$ the high-curvature phase.  Parameters for lipid-lipid interactions
are given for each pair of phases. } }
\end{table}
For the Langevin thermostat our base-line default parameters are given in Table
\ref{table:langevin_parameterization}.

\begin{table}[H]
\centering
\begin{tabular}{|lllll|}
  \rowcolor{atz_table1}
  \hline
$k_B T$        & $\nu_r$ & $\nu_n$     & $M$   & $I$   \\ \hline
0.23$\epsilon$ & 1$\tau$ & 3.333$\tau$ & 0.523$m$ & 0.523$m\sigma^2$ \\ \hline
\end{tabular}
\caption{\label{table:langevin_parameterization}{Parameters for the Langevin
thermostat. The values are expressed in terms of the
Lennard-Jones (LJ) characteristic scales.}}
\end{table}

\subsection{Vesicle Assembly and Equilibration}
{The vesicles were assembled by developing a geometric generator for spheres based on
Spherical Fibonacci Point Sets (SFPS)~\cite{Marques2013,Swinbank2006}.  As an
initial configuration, each bead was placed at the locations of the SFPS with the
bead orientation matching the outward normal of the sphere.  To equilibrate the
system, the vesicle was treated initially as homogeneous and simulations were
run under
the Langevin thermostat over a long trajectory.  During these simulations, the
beads both diffused and mixed within the membrane structure, and in some instances
spontaneously ejected or inserted into the membrane surface.  After the
equilibration stage, simulation studies were performed to investigate more
complex phenomena.}

{To investigate phase separation, the system was started with 
an equilibrated homogeneous
vesicle.  Some of the identities of the beads were then changed 
to represent the
high-curvature phase.  This allowed the system to evolve from a stable
configuration under
the interaction potentials of the two-phase system discussed in
Section~\ref{sec:CG_model}.  The membrane structures remained stable
throughout all of the simulations.  It was found there was not significant 
loss of beads within the two-phase membrane structures, other than 
the explicit budding events.}

\section{{Mapping Particle Configurations to Continuum Fields: \\Spherical Harmonics Representations}}
\label{sec:spectral_analysis}
{To investigate the coarse-grained vesicle structures, we introduce
techniques to map collective particle configurations to a continuum description.
Our continuum representations allow for using approaches from differential
geometry and statistical mechanics to characterize the membrane shape and
mechanical properties.}

\begin{figure}[H] \centering
\includegraphics[width=0.9\columnwidth]{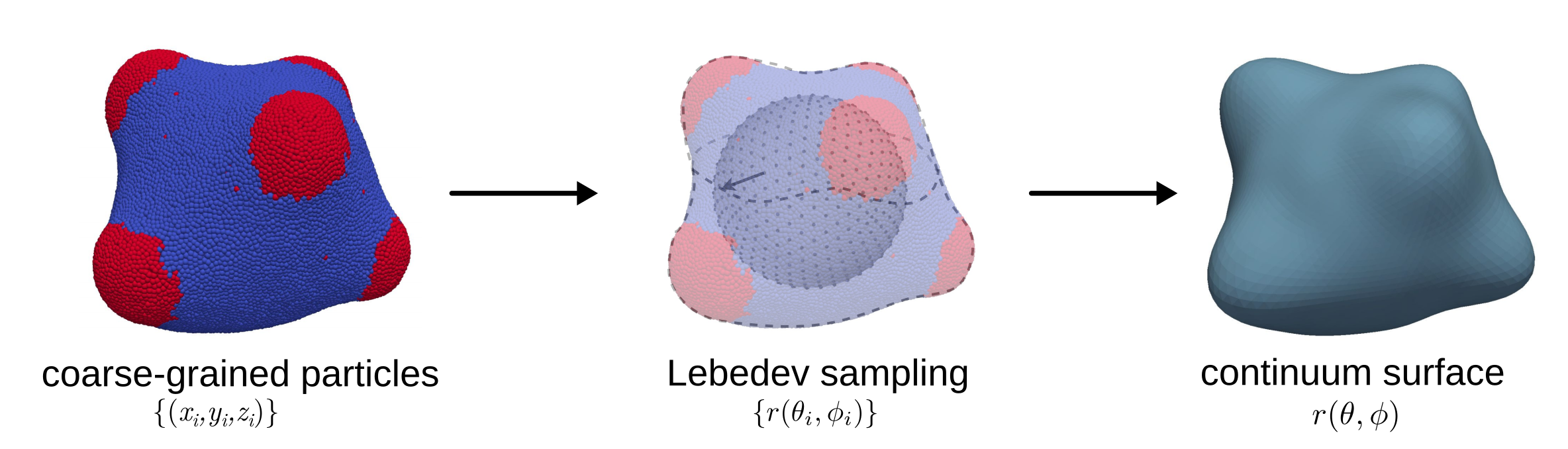}
\caption{{Mappings to the continuum representations.
The collective configurations of the particles $\{(x_i, y_i, z_i)\}$ are mapped
to continuum representations using spherical harmonics expansions for 
the radial shape functions $r(\theta,\phi)$ in equation \ref{sph_expansion}.  
Results are shown for $n=590$ Lebedev quadrature points for 
harmonics with degree $\ell \leq 21$.}} \label{fig:molecular_to_continuum}
\end{figure}

\subsection{{Continuum Surface Representations: Spherical Harmonics Expansions}}
\label{sec:surface_representation}
We represent the membrane surface in terms of the position function
$\boldsymbol{r}(\theta,\phi) = r(\theta,\phi) \Hat{\mathbf{e}}(\theta,\phi)
\sigma$. The $\theta$ denotes the polar angle and $\phi$ the azimuthal angle for
the spherical coordinate chart, $\Hat{\mathbf{e}}(\theta,\phi)$ denotes the
corresponding vector on the unit sphere, and $\sigma$ denotes our length-scale.
{The radial component $r$ of the function is expanded in spherical harmonics as}
\begin{eqnarray}
    r(\theta,\phi; \boldsymbol{a}) = \sum_{\ell,\hspace{0.1cm} |m|\leq \ell}
      a_{\ell m} Y_\ell^m(\theta,\phi),\;\;\;\; 
\label{sph_expansion}
    a_{\ell m} = \int_{\boldsymbol{r}} r(\theta,\phi)
      (Y_\ell^m(\theta,\phi))^* d\Omega.
\label{sph_expansion_L2}
\end{eqnarray}
{The $Y_{\ell}^m$ denotes the spherical harmonic of degree $\ell$ and order $m$.
The superscript $*$ denotes the complex conjugate.  The $a_{\ell m}$ and
expression in equation~\refp{sph_expansion_L2} gives the expansion coefficients for
$r(\theta,\phi)$ in spherical harmonics.  This can be viewed as the $L_2$-inner
product $a_{\ell m} = \langle r(\theta,\phi), Y_\ell^m(\theta,\phi)
\rangle_{L_2}$ over the sphere surface.}

{To approximate in practice the integration over the surface of the sphere
and the inner-product, we use Lebedev
quadratures~\cite{Lebedev1999,Lebedev1976}.  The Lebedev quadratures provide
weights $w_k$ and nodal locations $\mb{x}_k$ to sample scalar functions $f$ on
the spherical surface to approximate integrals as
\begin{eqnarray}
\int f(\mb{x}) d \Omega \sim \sum_{k=1}^m w_k f(\mb{x}_k).
\end{eqnarray}
{The Lebedev quadratures have a number of desirable properties that include a
high-order of accuracy and a more symmetric distribution of nodal samples than
uniform latitude-longitude based
methods~\cite{AtzbergerSoftMatter2016,AtzbergerGrossHydro2018,Lebedev1999}.} }

We estimate the radial function $r(\theta_k,\phi_k)$ as $\tilde{r}_k$ using the
coarse-grained beads comprising the membrane surface.  {This is computed
using $\tilde{r}_k$ of 
a ray generated by the angles $(\theta_k,\phi_k)$.  This is associated with the
$k^{th}$ Lebedev quadrature node on the unit sphere. The coordinates of the
closest bead to this ray is used on the surface.  These calculations used
$590$ Lebedev quadrature points to capture spherical harmonics of degree $\ell
\leq 21$.}
{The equation~\refp{sph_expansion_L2} is approximated by the discrete sum}
\begin{align}
\tilde{a}_{\ell m} &= \sum_k w_k \tilde{r}_k.
\label{sph_expansion_Lebedev}
\end{align}
The $w_k$ denotes the Lebedev quadrature weight associated with the $k^{th}$
node~\cite{Lebedev1999}.  Equation~\ref{sph_expansion_Lebedev} provides through
the spherical harmonics expansion a continuum representation of the membrane
surface geometry.

{Our particle-to-continuum mapping is demonstrated in 
Figure~\ref{fig:molecular_to_continuum}.  Here, the configuration is sampled
from a two-phase membrane and the shape is reconstructed using the continuum
representation based on spherical harmonics.  Our approaches also can be used
for the related problem of capturing the leading-order spherical harmonic modes
representing other fields and hydrodynamic flows on surfaces of vesicles,
see~\cite{AtzbergerSoftMatter2016,AtzbergerGrossMeshless2020,AtzbergerRower2022,
AtzbergerGrossSpectral2018,AtzbergerGrossHydro2018}.}

\subsection{{Bending Elasticity of Homogeneous Membranes}}
\label{sec:bending_elasticity_of_membranes}
We consider the elasticity theory for homogeneous membranes introduced by
Helfrich~\cite{Helfrich1973} and Canham~\cite{Canham1970} based on local mean
and spontaneous curvatures.  The free energy $E$ associated with membrane shape
is
\begin{align}
    E[\boldsymbol{r}] &= \int_{\boldsymbol{r}} \frac{k_c}{2}[2H(\theta,\phi;
      \boldsymbol{a}) + c_0]^2 dA + \lambda \int_{\boldsymbol{r}} dA.
\label{helfrich_energy}
\end{align}
{The $k_c$ denotes the bending rigidity, $H$ the mean curvature, $c_0$ the
spontaneous curvature, $dA$ the infinitesimal vesicle surface area, and $\lambda$
the tensile stress, serving as a Lagrange multiplier to maintain constant area.}

{We linearize the free energy in equation~\refp{helfrich_energy} in the case of
vanishing spontaneous curvature $c_0 = 0$ around the reference configuration of
a sphere and expand in spherical harmonics.  In this case, the tensile stress
term does not play a role in the second variation~\cite{Helfrich1989}. The free
energy can be expressed in terms of spherical harmonics as}
\begin{align}
  \tilde{E}[\mb{r}] = \tilde{E}[\{a_{\ell m}\}] 
  = \frac{k_c}{2r_0^2} \sum_{\ell,m} |a_{\ell m}|^2
    \left[\ell(\ell+2)(\ell^2-1)\right].
\label{helfrich_second_var}
\end{align}
The $a_{\ell m}$ denote the spherical harmonic expansion coefficients of the
shape discussed in Section \ref{sec:surface_representation}.

For physical vesicles comprised of molecular or particle constituents, the
thermal undulations are modeled in equation~\refp{helfrich_second_var}
and~\cite{HelfrichSizeDep1986} only up to the leading spherical harmonic modes
above the molecular length-scales.  Since the material is not a pure continuum,
this results in aliasing artifacts in the harmonics description and for
sufficiently small vesicles can result in what manifests as effective
enhancement of the amplitudes of the fluctuations of harmonic modes.

Using the linearized theory of equation~\refp{helfrich_second_var} to estimate a
bending rigidity $k_c$ can then lead to an under-estimate of the mechanical
rigidity.  Helfrich derived correction terms to account for higher-order
contributions in~\cite{HelfrichSizeDep1986} giving the size-dependent effective
bending rigidity
\begin{equation}
k_c' = k_c - \frac{k_B T}{8 \pi} \log{M}.
\label{helfrich_effective_rigidity}
\end{equation}
For spherical geometries, the $M$ represents half the number of amphilic
molecules of a lipid bilayer membrane or more generally the number $N$ of
particles comprising a leaflet of the membrane surface.  The term involving $M$
can be viewed as an entropic contribution capturing neglected degrees of freedom
of the system in the linearized elasticity theory such as higher-order
contributions in the free energy~\cite{HelfrichSizeDep1986}.  The molecular
parameter $M = N \approx \kappa 4 \pi r_0^2$ is proportional to the area of the
membrane surface.  {The equation~\refp{helfrich_effective_rigidity} can be expressed as}
\begin{align}
\beta k_c' &= K - C \log{r_0},
\label{effective_rigidity_and_radius_KC}
\end{align}
where $\beta^{-1} = k_B T$. Within Helfrich's theory for the area corrections of the linearized elasticity
when treating vesicles as quasi-spherical~\cite{HelfrichSizeDep1986}, the
coefficients would be $K = \beta k_c - \frac{1}{8 \pi} \log{(\kappa 4 \pi)}$ and
$C = {1}/{4 \pi}$.

We expect more generally for other entropic effects to contribute to the elastic
bending modulus with a similar scaling as
equation~\refp{effective_rigidity_and_radius_KC}.  This would correspond to other
values of $K$ and $C$.  We show in Section~\ref{sec:size_dep_Helfrich} that such
a scaling theory can be used with fit values of $K$ and $C$ to characterize how
the estimated bending rigidity varies with the size of our coarse-grained
vesicles.

\subsubsection{Spherical Harmonics Conventions and Scale Separation}
\label{sec:sph_harmonics_conv}
In practice, we have found it convenient to perform analysis using a real-valued
spherical harmonics basis $X_\ell^m$, $Z_\ell^m$ expressed as $Y_\ell^m =
X_\ell^m + iZ_\ell^m$ with $a_{\ell m} = \frac{1}{2}(x_{\ell m}-iz_{\ell m})$
for $m > 0$ and $x_{\ell m}, z_{\ell m}\in \mathbb{R}$.  Since the membrane
surface function is always real-valued, we have in the spherical harmonics
representation that $a_{\ell m} = a_{\ell m}^*$, which requires that
$x_{\ell m} = x_{l-m}$ and $z_{\ell m} = -z_{\ell-m}$.  This allows us to
express the radial function as
\begin{align}
\label{equ:real_sph_harm_expansion}
    r(\theta,\phi) &= \sum_{\ell=0}^{\infty}\sum_{m:-\ell \leq m \leq \ell}
      a_{\ell m} Y_\ell^m  
	  = \sum_{\ell=0}^{\infty} \big[a_{\ell,0} Y_\ell^0 +
      \mkern-15mu\sum_{m:1 \leq m \leq \ell} x_{\ell m}X_\ell^m +
      z_{\ell m}Z_\ell^m\big].
\end{align}
This yields a real-valued basis comprised of $Y_\ell^m$ for $m = 0$ and
$X_\ell^m$, $Z_\ell^m$ for $\ell > 0$, $m > 0$.  This has corresponding
expansion coefficients $a_{\ell,0}$, $x_{\ell m}$ and $z_{\ell m}$ similar
to~\cite{AtzbergerGrossHydro2018,AtzbergerGrossSpectral2018}. {This can
be related to our spherical harmonics expansion coefficients $a_{\ell m}$ by}
\begin{equation}
    a_{\ell m} =
    \begin{dcases}
        \frac{1}{2}(x_{\ell m}-iz_{\ell m}),     & m > 0 \\
        a_{\ell,0},                          & m = 0 \\
        \frac{1}{2}(x_{\ell,|m|}-iz_{\ell,|m|}), & m < 0.
    \end{dcases}
\end{equation}

{When considering the continuum mechanics which arises from the collective
mechanics associated with the molecular interactions, it is important to
consider the length-scales associated with the observed responses.  This is
especially important when length-scales approach molecular scales reaching the
limits of a purely continuum interpretation.  In the spectral analysis this
corresponds to considering the length-scales associated with the different modal
responses.   To help with such interpretations, we develop relations between the
key parameters to characterize spherical harmonics expansions and related
spectral analysis.}

{Consider the characteristic length-scale of the particle/molecular 
interactions, which we denote by $d$.  For our coarse-grained vesicle
models, this is taken to be the bead
size $d = \sigma$.  For spherical harmonics expansions, this gives the critical
wave-length associated with continuum responses of the system.
One can think of the
length $d$ as that of a small arc that is drawn along the equator of a sphere of
radius $r_0$.  For a quasi-spherical vesicle, the spherical harmonics
expansion of equation~\refp{sph_expansion_L2} can be used to estimate the radius as $r_0 =
{a_{00}}/{\sqrt{4\pi}}$. The $a_{0 0}$ is the coefficient of the constant
mode.  In spherical coordinates, the azimuthal angle is given by $\phi = d/r_0$.
The spherical harmonic $Y_\ell^m$ has azimuthal angle dependence through the
term $e^{im\phi}$.  This has the largest spatial frequency when $m =\pm\ell$,
giving the smallest wave-length resolved as $\lambda_\ell \sim r_0/m \sim
r_0/\ell$.}

{In practice, for correspondence with continuum mechanics we should consider
modes $\ell$ with $\lambda_\ell \gtrsim d$.  When $d = \sigma$ is the bead size,
we see this scaling analysis indicates that the local molecular or
particle-level contributions start to dominate the spherical harmonics expansion
when the harmonic degree $\ell$ is on the scale $\ell \sim r_0/d \sim
r_0/\sigma$.  This suggests the modes with $\ell \lesssim r_0/\sigma$ capture
most significantly the responses of the system at the level of continuum
mechanics.  For larger degrees $\ell \gtrsim r_0/\sigma $, the modal responses
depend more directly on signatures of the local molecular level interactions and
noise.  To relate the exhibited behaviors of our vesicles to continuum mechanics
we shall focus on modal responses with $\ell \lesssim r_0/\sigma$. }

\subsubsection{{Spectral Analysis of Passive Shape Fluctuations and Bending Elasticity}}
\label{sec:passive_shape_fluctuation_estimator}
We use the passive thermal fluctuations of the vesicle shape to obtain
information about the elastic bending modulus.  From equilibrium statistical
mechanics,  the shape fluctuations are governed by the Gibbs-Boltzmann
distribution
\begin{align}
\rho[\{a_{\ell m}\}] = \frac{1}{Z} \exp\left(-E[\{a_{\ell m}\}]/k_B{T}  \right),
\end{align}
where $Z$ represents the canonical partition function. Using the free energy of 
equation~\refp{helfrich_second_var}, the shape modes of 
$\rho[\{a_{\ell m}\}]$ have a Gaussian distribution with mean zero and variance
\begin{align}
\mbox{Var}[a_{\ell m}] = \langle a_{\ell m}^2 \rangle =
  \frac{r_0^2}{\beta k_c}\left[\ell(\ell+2)(\ell^2-1)\right]^{-1}.
\label{equ:var_modes}
\end{align}
{The covariance between different modes is zero with
$\mbox{Cov}[a_{\ell m},a_{\ell' m'}] = \langle a_{\ell m} a_{\ell' m'}
\rangle = 0$, where $\ell \neq \ell'$, $m \neq m'$.}  As $\ell$ becomes large, the
variance scales as $\mbox{Var}[a_{\ell m}] \sim \ell^{-4}$.  {The 
equation~\refp{equ:var_modes} can be used to estimate the bending 
rigidity $k_c$ from the passive fluctuations of the vesicle.}

In practice, we use the real-valued basis expansion of
equation~\refp{equ:real_sph_harm_expansion}.  We track $\text{Var}[a_{\ell,0}]$,
$\text{Var}[x_{\ell m}]$ and $\text{Var}[z_{\ell m}]$ with $\ell > 0$, $m > 0$.
{When $\text{Cov}[x_{\ell m},z_{\ell m}] = 0$, the
variance can be expressed as}
\begin{align}
  \text{Var}[a_{\ell m}] =
  \begin{dcases}
    \text{Var}[a_{\ell, 0}], & m = 0 \\
    \frac{1}{4}(\text{Var}[x_{\ell,|m|}]+\text{Var}[z_{\ell,|m|}]), & m \neq 0.
  \end{dcases}
\end{align}
{When the variances are uncoupled for the different harmonic orders $m$, as in
the elasticity theory of Section \ref{sec:bending_elasticity_of_membranes}, 
averaging can be used to obtain}
\begin{equation}
  \langle a_{\ell}^2 \rangle := \frac{1}{2\ell+1}\sum_{|m|\leq \ell}
    \langle a_{\ell m}^2 \rangle.
\label{helfrich_theory_variance}
\end{equation}
For spherical geometries in the basis expansion of Section
\ref{sec:surface_representation}, the mean is zero
$\langle a_{\ell m} \rangle = 0$ for $\ell \neq 0$.
{The bending rigidity $k_c$ is estimated by using a linear regression in 
$\log$-space with the estimator}
\begin{equation}
\log{\langle a_{\ell}^2 \rangle} = b-m\log{\left[\ell(\ell+2)(\ell^2-1)\right]}.
\label{rigitidy_regression_eq}
\end{equation}
The $b = \log{(2 r_0^2/\beta k_c)}$ and $m = 1$.  In practice, 
the effective bending rigidity $k_c'$ is estimated from
equation~\refp{effective_rigidity_and_radius_KC}.  {In
Section~\ref{sec:size_dep_Helfrich}, we showed that the empirical data exhibits the
expected logarithmic scaling for the parameter values of $K$ and $C$ 
obtained from fitting simulations varying the vesicle size $r_0$.}

\section{Results} \label{sec:results}

\subsection{Phase Separation in Heterogeneous Vesicles}
\label{sec:results_phase_sep}

\begin{figure}[H] \centering
\includegraphics[width=0.4\columnwidth]{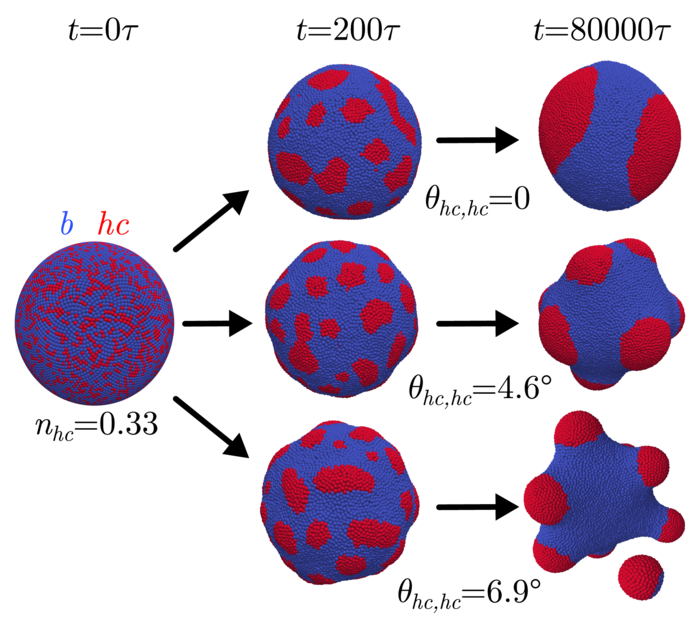}
\caption{{The preferred curvatures and phase separation
are shown for a few cases.  The phase preferred curvatures 
can impact coarsening and the vesicle shape.  The top has matched
curvatures between phases with $\theta_{hc,hc} \sim 0$ and the middle 
has intermediate differences.  The bottom shows large differences where 
domains can form buds and 
daughter vesicles.
The simulations consisted of $8000$ beads with $n_{hc} = 0.33$, $k_b T =
0.20\epsilon$, $\mu_{b,b}=3.0$, and $\mu_{b,hc}=6.0$.}}
\label{fig:phase_sep_phenom} \end{figure}

{The heterogeneity of biological membranes and synthetic soft materials 
can play a significant role in shaping the geometry and in 
mechanical responses~\cite{Baumgart2003, SorreLipidSorting2009,
Lingwood2010, KimZasadzinskiSquires2011, DogicColloidalMembraneRafts2014}}. 
{For both homogeneous and heterogeneous vesicles, we study how the mechanics
depends on the concentration ratio and preferred curvatures of the
phases.  A few behaviors exhibited in our simulations are shown in 
Figure~\ref{fig:phase_sep_phenom}.}

{In our models the phase separation is driven primarily 
by the different species affinities, even when 
the preferred curvatures are the same.   As the
preferred curvature of the second phase is increased, it is found 
that the coarsening dynamics can stall given competition with curvature effects.  
The vesicles appear to 
exhibit meta-stable domains related to the elastic energy
associated with the formation of bulged phase-separated domains.  This
results in a bending elasticity for the vesicle which presents a
sufficiently large energy barrier inhibiting domains 
from further growth and merging during fluctuations.  As the preferred
curvature increases further, the coarsening dynamics were found to 
proceeds until a critical size, after which daughter vesicles bud 
from the vesicle.  A few instances of these behaviors seen in our 
simulations are shown in Figure~\ref{fig:phase_sep_phenom}. }

{In the case of a significant line tension, as opposed to a
preferred curvature contrast, there can still be formation of buds or highly
curved sub-domains.  The parametrizations are considered where the line tensions
are not sufficiently large to overcome the local bending energy.  While the
particle interactions are related in the coarse-grained potential, it is the
local preferred curvature terms in the energy that drive the resulting geometry
of our vesicles.  For spherical topologies, the
phase-separation will tend to proceed with the formation of two meta-stable
domains at polar regions with the base phase in-between.  This results in a large
energy barrier inhibiting further merging into a single domain.
In the intermediate regime with $\theta_{hc,hc} \sim 4.6$, the preferred curvature
difference stalls the coarsening to yield many
distinct sub-domains.
We investigate both the passive shape fluctuations and mechanical responses to 
active deformations in Section~\ref{sec:size_dep_Helfrich}--~\ref{sec:transport_experiment}.
}

\subsection{Homogeneous Vesicles and Size-Dependence of the Bending Elasticity}
\label{sec:size_dep_Helfrich}

{As a baseline for our studies, we first consider for homogeneous vesicles
how their passive shape fluctuations and mechanical responses depend on the
vesicle size.}

\begin{figure}[H] \centering
\includegraphics[width=0.45\columnwidth]{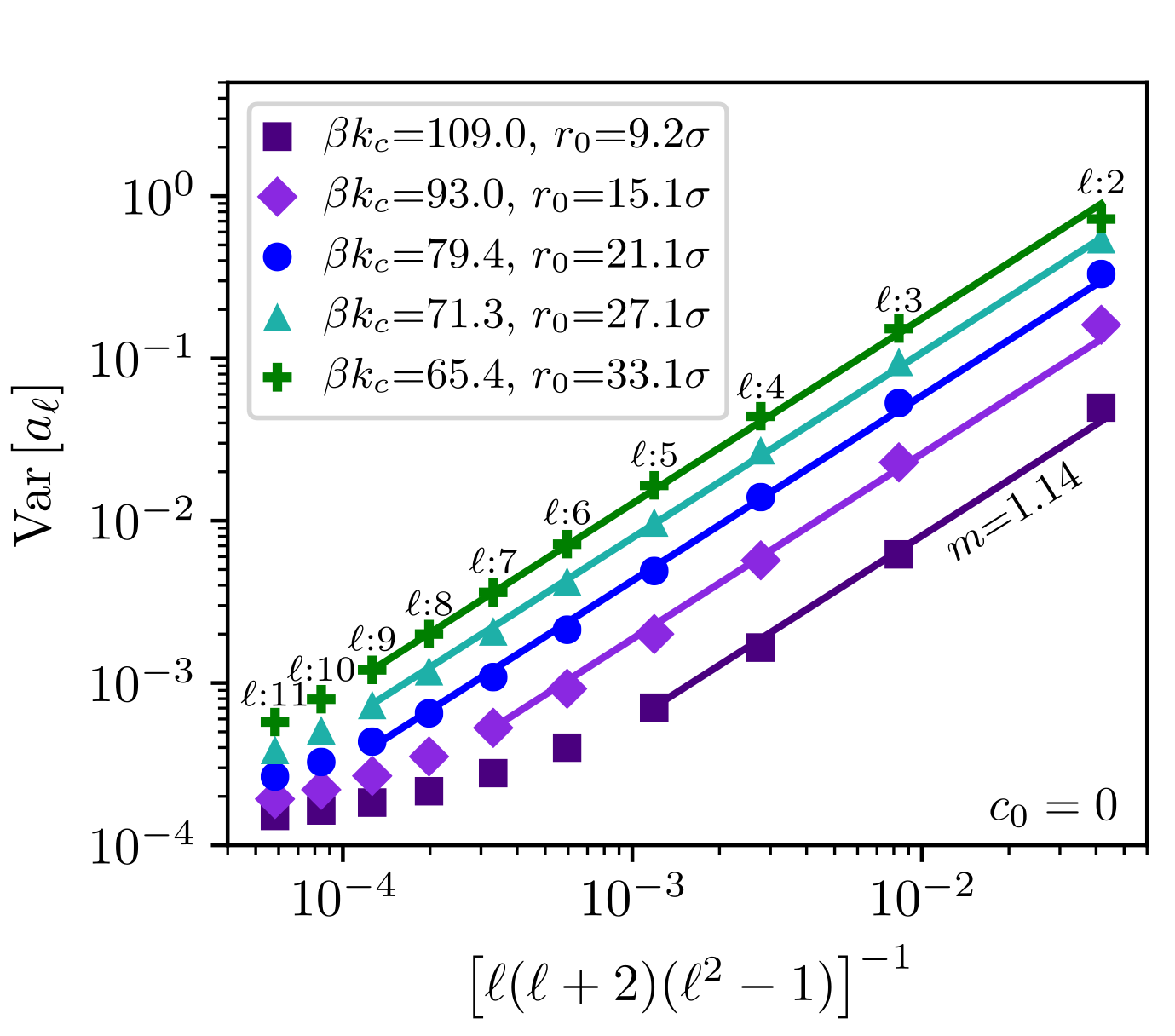}
\caption{{The fluctuation spectrum of vesicles of different sizes.  For
passive shape fluctuations mapped to harmonic modes of degree $\ell$,
the variances 
are in good agreement with the continuum elasticity theory in equations\refp{helfrich_effective_rigidity}, 
\ref{helfrich_theory_variance},
and predictions by Helfrich~\cite{Helfrich1989}.  Results are for 
$1000$ time-samples averaged over $5$ runs for each 
vesicle size.}}
\label{fig:rigidity_verses_radius_spectrum}
\end{figure}

In Helfrich~\cite{HelfrichSizeDep1986}, theory was developed for small vesicles
predicting that the effective bending elasticity observed from shape
fluctuations can depend significantly on the vesicle size.  The coarse-grained
models are used to further investigate the role of the vesicle size.  We
investigate the bending elasticity of homogeneous vesicles based on passive
shape fluctuations using the methods developed in
Section~\ref{sec:passive_shape_fluctuation_estimator}. {The mechanical
responses are considered over the range of sizes from $1255$ beads with $r_0
\sim 9\sigma$ to $20088$ beads with $r_0 \sim 36\sigma$. The surface
area fluctuations are found to be less than $0.05\%$ for all sizes.
For our coarse-grained vesicles, we show the scaling of the variances of
the spherical harmonic modal responses $a_{\ell m}$ with degree $\ell$ in
Figure~\ref{fig:rigidity_verses_radius_spectrum}.}  

\noindent
{The modes are averages over 
order $m$ as presented in equation~\refp{helfrich_theory_variance}.  The linearized
continuum elasticity theory predicts a scaling of $\mbox{Var}[a_{\ell m}] \sim
\left[\ell(\ell +2)(\ell^2 -1)\right]^{-1}$. While holding the particle
interaction parameters fixed for the coarse-grained model, we find the estimated
elastic bending modulus decreases as the size of the vesicle increases, see
Figure~\ref{fig:rigidity_verses_radius_spectrum}.}

\begin{figure}[H] \centering
\includegraphics[width=0.45\columnwidth]{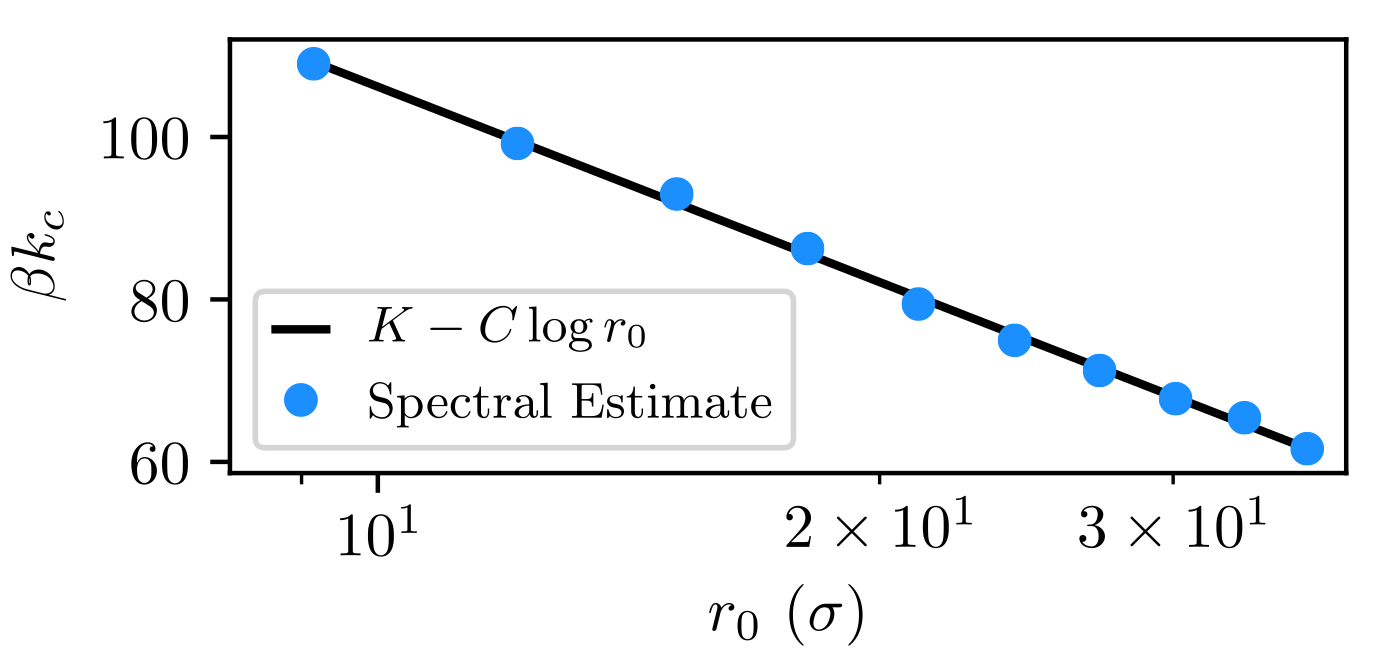}
\caption{{The bending elasticity for vesicles of different sizes.
Shown is the elastic bending modulus exhibiting logarithmic scaling similar to
the correction terms derived in Helfrich~\cite{Helfrich1989} and equation
\ref{helfrich_effective_rigidity}.  The fitting parameters are $K = 186.08$
and $C = 34.70$.}}
\label{fig:rigidity_verses_radius} 
\end{figure}

{For the linearized elasticity theory,  entropy correction terms were derived 
for estimating 
the bending rigidity when varying the vesicle 
size~\cite{HelfrichSizeDep1986}.  The correction terms account 
for entropic contributions neglected in linear expansions.  
Our results show a similar trend with our
coarse-grained vesicles having mechanical responses that exhibit a scaling
similar to equation~\refp{effective_rigidity_and_radius_KC}.} {The estimated
bending modulus is shown in log-space as the vesicle size is varied, see
Figure~\ref{fig:rigidity_verses_radius}.}  We find a good fit is obtained for the
scaling theory with the regression parameters $K = 186.08$ and $C = 34.70$ in
equation~\refp{effective_rigidity_and_radius_KC}.

\subsection{Shape Fluctuations of Heterogeneous Vesicles}
\label{sec:two_point_corr}

{Heterogeneous vesicles are found to exhibit interesting spatial-correlations
associated with sub-domain mechanics. This couples modes and 
makes traditional spectral analysis challenging.  As an alternative we make
comparisons using real-space two-point correlation functions of the undulations 
of the membrane surface for heterogeneous and 
homogeneous vesicles.  The case is considered where 
heterogeneity occurs at scales comparable to the vesicle size. The shape fluctuations 
are characterized by the two-point correlations $\psi$ associated with the
radial shape function $r(\theta,\phi)$.}

\begin{figure}[H] \centering
\includegraphics[width=0.8\columnwidth]{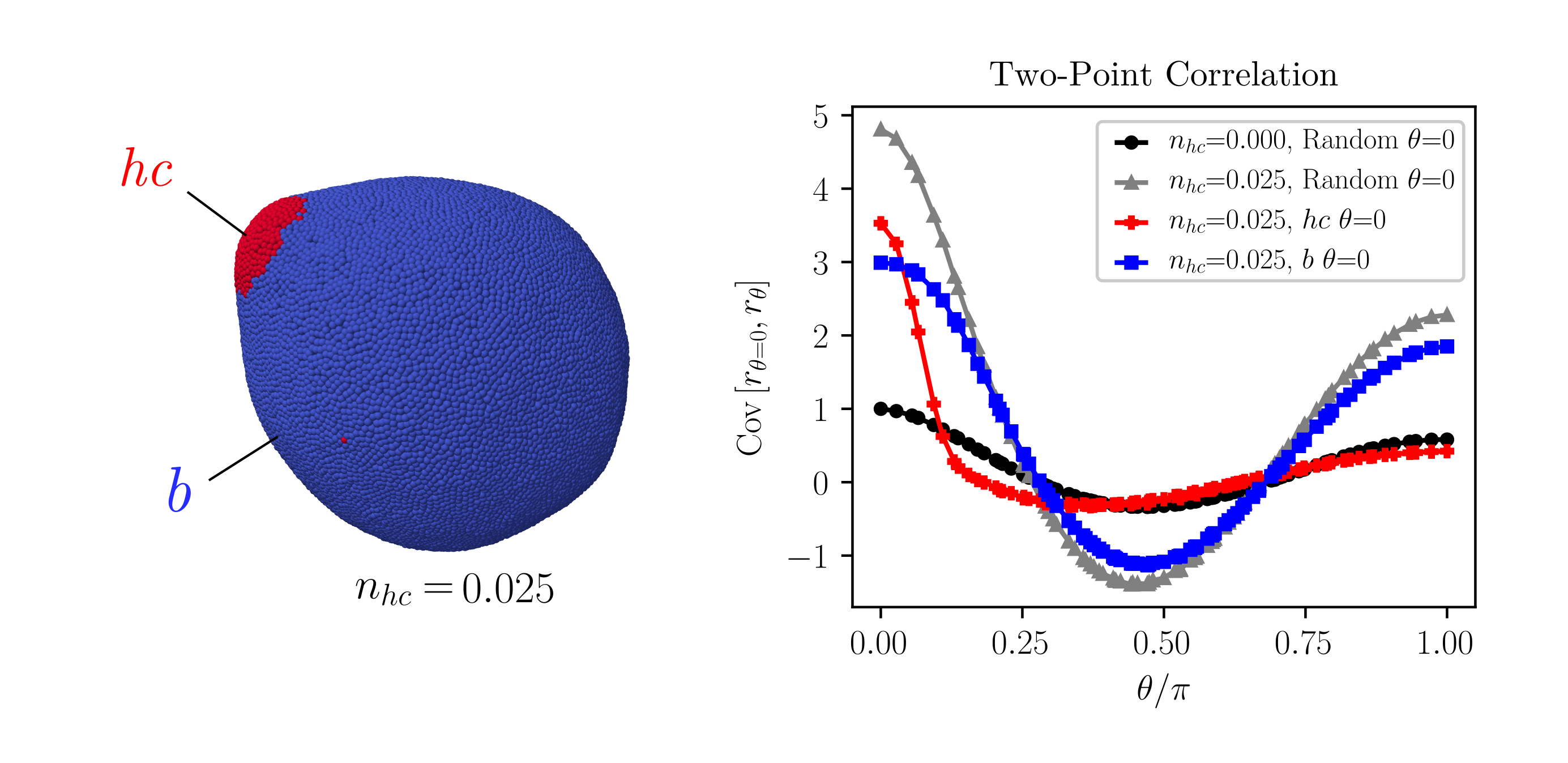}
\caption{{The surface two-point correlations are
shown for the radial shape function
$r(\theta,\phi)$ for homogeneous $n_{hc} =0$ and heterogeneous vesicles $n_{hc}
> 0$.  The cases correspond to when the base-point is chosen (i) as any location at random on
the surface, denoted ``Random $\theta=0$", (ii) 
to be within a high-curvature domain of phase $hc$, denoted ``$hc\,\, \theta=0$", 
and (iii) to be any point of the base phase $b$, denoted by ``$b\,\, \theta=0$".
The base-point is taken to have coordinate $\theta = 0$.
Sampling was performed for $5$ random
base-beads for 100 time-samples averaged over $5$ simulations runs.  Results
are normalized in scale by the variance at $\theta = 0$ of the homogeneous
vesicle.  }}
\label{fig:two_pt_corr_verses_polar_angle} 
\end{figure}

{The correlations are computed by choosing a base point for which the 
radial component is correlated with other points on the vesicle surface.
There are three cases for choosing base-points, (i) as any location at random on
the surface, denoted ``Random $\theta=0$", (ii) 
to be within a high-curvature domain of phase $hc$, denoted ``$hc\,\, \theta=0$", 
and (iii) to be any point of the base phase $b$, denoted by ``$b\,\, \theta=0$".
The base-point is taken to have coordinate $\theta = 0$.  The case (i) is used 
primarily for homogeneous vesicles.  The cases (i) and (ii) is used for
the heterogeneous vesicles.  The main difference is in case (ii) the base point
is chosen within one of the phase-separated domains.  This captures fluctuations
within and in the vicinity of the domains of the vesicle.  The results
are normalized in scale by the variance at $\theta = 0$ of the homogeneous
vesicle.}

{In more detail, to obtain $\psi$ the chosen base-point can be thought of 
as being at the north pole of the vesicle with angles $(\theta_0,\phi_0) = (0,0)$.
In practice, this is achieved by applying a rotation to the 
vesicle to transform it into this standard orientation.  
The base-point radial coordinate is then correlated with the other
points of the vesicle at the polar angles $\theta$ by averaging the results over the
azimuthal angle $\phi$.  For the given case (i)--(iii) of base-point, this gives
the two-point correlation function $\psi = \mbox{Cov}[r_{\theta=0}r_\theta]$. In
practice, we approximate this by the estimator $\tilde{\psi} = 1/N\sum_{i=1}^N
r_{\theta=0}^{[i]} r_\theta^{[i]} - \bar{r}_{\theta=0}\bar{r}_\theta$.  The
$r_{\theta=0}^{[i]}$ is the $i^{th}$ sample of the base-point.  The
$r_\theta^{[i]}$ is the $i^{th}$ sample of the radius for any point with polar
angle $\theta$.  This is averaged over a sampling of base points consistent with
one of the cases (i)--(iii).  The $\bar{r}_{\theta}$ denotes the empirical 
mean of the radius of the points with polar angle $\theta$.  
Results for homogeneous vesicles with $n_{hc}=0.0$ and
heterogeneous vesicles with $n_{hc}=0.025$ are shown in
Figure~\ref{fig:two_pt_corr_verses_polar_angle}.}

{It is found the two-point correlations for case (i) for both homogeneous and heterogeneous vesicles exhibit a strong negative correlation at around $\theta = \pi/2$.  
A strong positive correlation is also found at around $\theta = \pi$.  
These angles and correlations indicate significant ellipsoidal shape fluctuations in
the overall geometry of the vesicles.  This behavior also is seen to persist when also
considering correlations at a base-point chosen in the base phase $b$, case (iii).  
When the base-points are chosen within the high-curvature domain of phase $hc$ (case (ii)), 
more complicated behaviors arise.  The base-point plays an important role, since within
the $hc$ regions it is sensitive to the local mechanics of the phase-separated domains.
For the base-points in the $b$ regions characterizes locally the base-phase with more
distant coupling to the phase domain mechanics.  The 
correlations appear to be strongest within the
high-curvature domain with relatively weak correlations with the regions of the
base phase $b$, see Figure~\ref{fig:two_pt_corr_verses_polar_angle}.  The 
correlation functions are normalized by the $n_{hc}=0.0$ surface variance at
$\theta=0$, $\mbox{Var}[r_{\theta=0}] = 0.38$.}

{Our results for two-point correlations show some of the 
challenges inherent in developing estimators for the bending elasticity
from a spectral analysis of shape fluctuations of heterogeneous vesicles.  
The heterogeneity
breaks spatial symmetries resulting in coupling between spectral modes
posing challenges for theory and analysis.  Also, the phase domains
can be of a comparable scale to the vesicle posing further issues.
The phase domains have different local mechanical properties and can diffuse
during fluctuations, undergo rearrangements, or merge.  As an alterative 
for heterogeneous vesicles, we perform further 
simulations to actively drive deformations of the vesicles
to better understand their mechanics and the roles of the phase domains.}

\subsection{Compression of Heterogeneous Vesicles: Mechanical Responses}
\label{sec:compression_experiment}

{In the molecular biology of cells, a central challenge is to understand
the mechanisms by which cells sense and transduce mechanical stimuli into
biochemical signals~\cite{Janmey2007,Mogilner2018,Alberts2002}.  This includes
both passive and active mechanical responses of cellular and sub-cellular
structures~\cite{Mogilner2003,Hoffman2009,Boal2002,Kirchhausen2000}.  The
coupling of chemical kinetics and mechanics is also important in the
self-assembly and mechanical responses of synthetic soft
materials~\cite{Fery2007,Hamley2003,Boal2002,
DogicColloidalMembraneRafts2014}.}  With the aim of understanding general
principles, laboratory experiments and simulations have been performed with
model physical systems. In~\cite{Schafer2013}, giant liposomes have been
compressed between parallel plates to investigate force compression curves of
the lipid membrane and in the presence or absence of an actin cortex.
In~\cite{Barlow2016}, coarse-grained studies were performed for the relaxation
times of homogeneous vesicles compressed by Atomic Force Microscopic (AFM) to
obtain dependence on the applied force of the time-scale for stress relaxation.

\begin{figure}[H] \centering
\includegraphics[width=0.99\columnwidth]{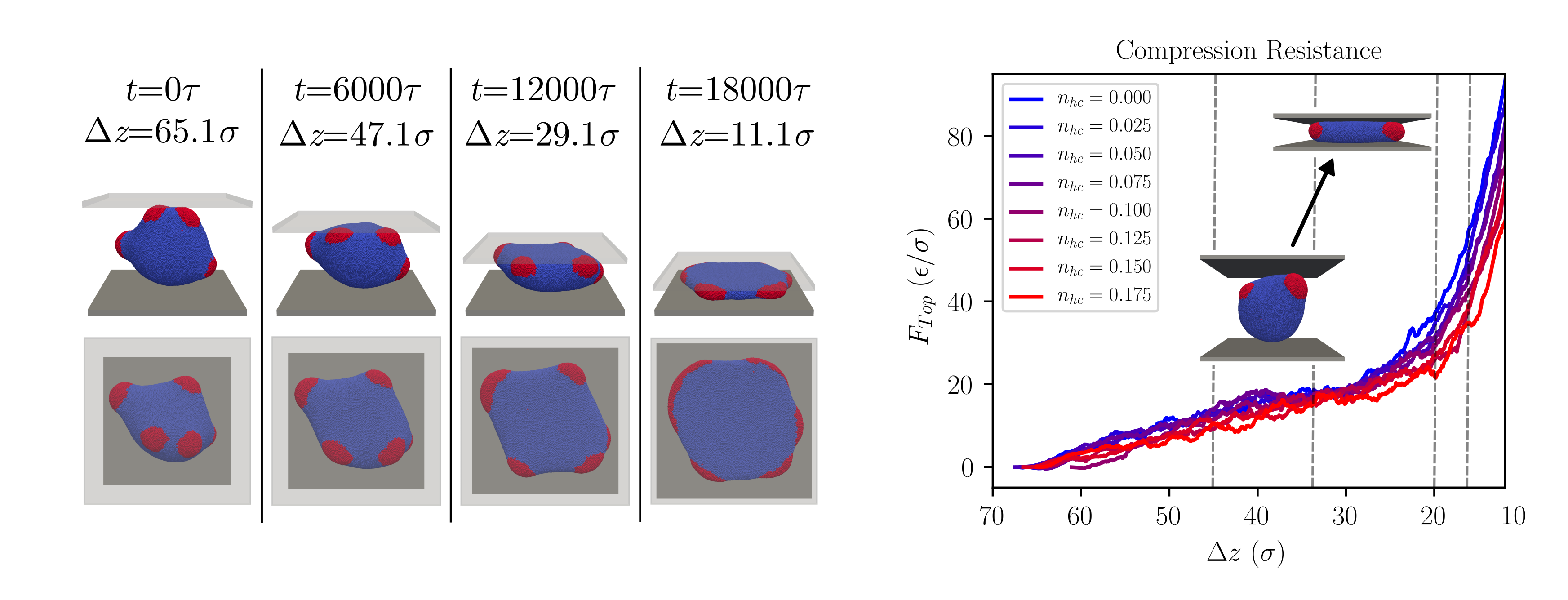}
\caption{{The compression forces of the deforming heterogeneous vesicles.  A 
sample multi-phase vesicle is shown during different stages of compression when $n_{hc} =
0.175$ \textit{(left)}.  For vesicles with different phase mixtures $n_{hc}$,
the resisting forces are shown \textit{(right)}.  For the dashed lines 
at compressions $\Delta{z} =
45\sigma,\Delta{z} = 33\sigma$, $\Delta{z} = 20\sigma$, and $\Delta{z} =
16\sigma$, the force is compared for the different phase concentrations 
in Figure~\ref{fig:max_force_Top_verses_nhc}.}}
\label{fig:squish_phenomenology}\label{fig:force_Top_verses_wall_sep}
\end{figure}

{We perform studies to probe the mechanics of heterogeneous vesicles
by compressing them between flat plates. A collection of
representative meta-stable coarse-grained vesicles are used 
having varying levels of concentration for 
a high-curvature species.  Vesicles of the type discussed in
Section~\ref{sec:results_phase_sep} are used
with $\theta_{hc,hc} = 5.73^{\circ}$.
The specific concentrations considered are
$n_{hc} = 0.025 k$, where $k=0,1,2,...,7$.}  The $n_{hc}$ gives the percentage of
high-curvature species in the membrane.  {Our equilibrated non-spherical vesicles
have a size characterized by the expansion coefficient $a_{00}$ for the
harmonic mode $Y_0^0$ as in Section~\ref{sec:sph_harmonics_conv}.}  Our vesicles
have similar characteristic radius $r_0$ ranging from $r_0 \sim 30.1\sigma$ for
$n_{hc} = 0.0$ to $r_0 \sim 28.9\sigma$ for $n_{hc} = 0.175$. 
{Vesicles are compressed by moving a top wall downward toward a stationary wall
below. The top wall is moved at a constant speed of $v = 0.003 \sigma/\tau$.
The steric particle-wall interactions are modeled by a 9-3 LJ-potential with
depth $\epsilon_{\text{wall}} = 0.01\epsilon$ with length-scale
$\sigma_{\text{wall}} = \sigma$.  The particle-wall force is computed using the
approach in Appendix~\ref{appendix:compression}.}

\begin{figure}[H] \centering
\includegraphics[width=0.5\columnwidth]{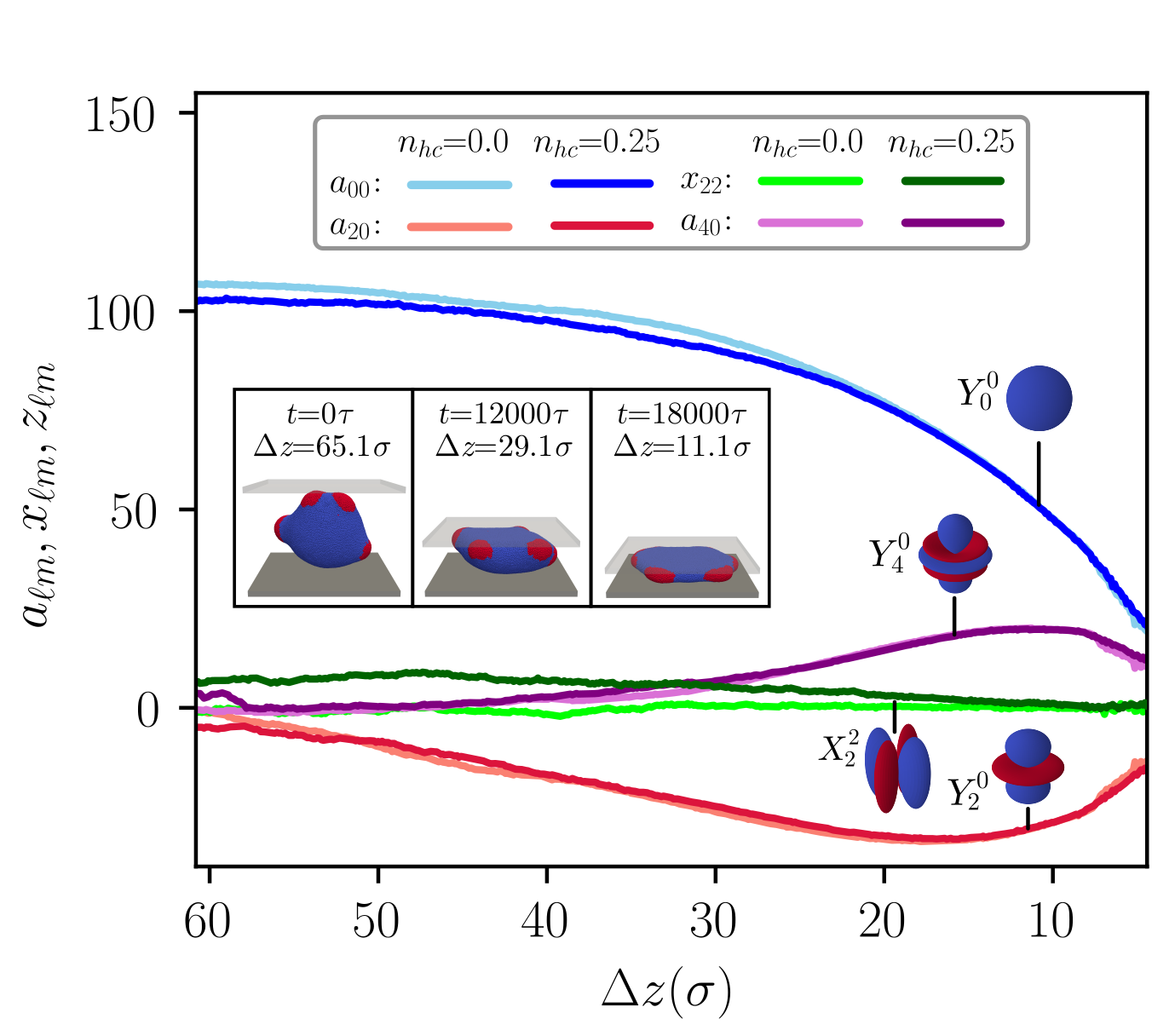}
\caption{{The shape changes and domain rearrangements during compression.
For vesicles with $n_{hc} =0$ and $n_{hc} = 0.175$, the shape changes are
characterized by select spherical harmonic modes with coefficients $a_{00}$,
$a_{20}$, $x_{22}$, and $a_{40}$. These correspond to the real-valued
harmonics shown as radial shapes with blue for positive and red for negative
values for $Y_0^0$, $Y_2^0$, $X_2^2$, and $Y_4^0$. 
}}
\label{fig:compression_shape_transformation}
\end{figure}

\begin{figure}[H] \centering
\includegraphics[width=0.45\columnwidth]{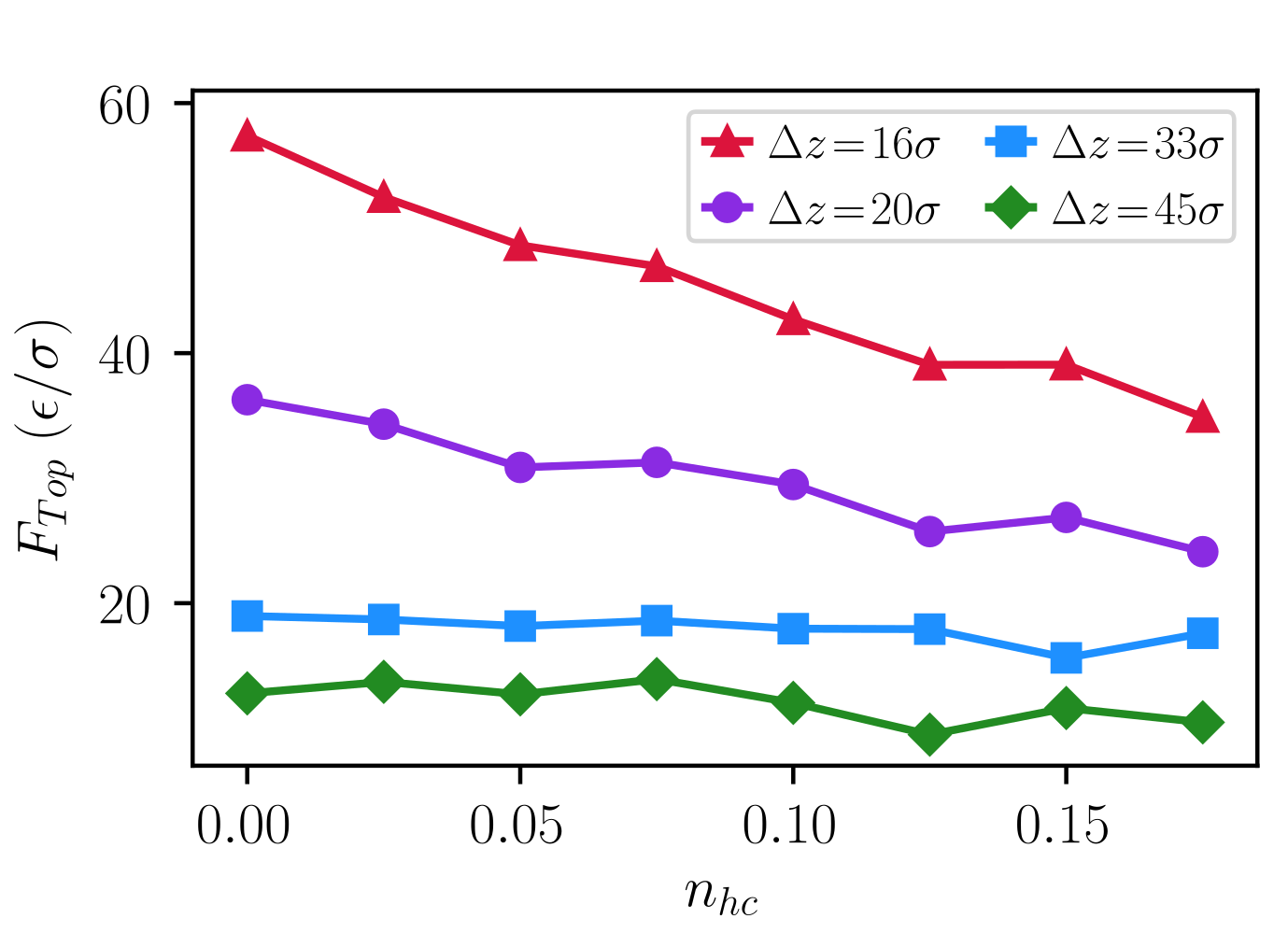}
\caption{{The compression forces of the deforming heterogeneous vesicles. Shown is
the 
$z$-component of the resistance force when varying the concentration $n_{hc}$.  
The $\Delta{z}$ indicates the wall-separation distance.}} 
\label{fig:max_force_Top_verses_nhc}
\end{figure}

{During compression, the high-curvature phase domains rearrange
to be parallel to the walls around the free perimeter of the flattening vesicle, 
see Figure~\ref{fig:squish_phenomenology}. For heterogeneous structures arising in
cell biology, the rearrangement of such protruding domains could provide
potential mechanisms for mechanosensing large compressive deformations. 
During most of the compression, the homogeneous and heterogeneous
vesicles take on overall ellipsoidal-like shape with similar $Y$ modes.
However, the shape mode $x_{22}$ differs significantly for the heterogeneous 
case given the role of the phase domains, see
Figure~\ref{fig:compression_shape_transformation}.  While the overall shapes
following similar trends, the heterogeneous vesicles can accommodate better the
stresses associated with the large deformations by rearrangement of the
high-curvature preferring domains towards the areas of larger curvature. 
This results in a lower energetic cost for the deformation and smaller
resistance forces since the high-curvature domains rearrange to occupy 
bending regions at their preferred curvatures, see Figure~\ref{fig:max_force_Top_verses_nhc}
and Figure~\ref{fig:squish_phenomenology}.
}

We investigate the resisting forces the vesicles exert on the walls as the level
of heterogeneity increases in Figure~\ref{fig:max_force_Top_verses_nhc}.  {It is 
found the resisting forces the vesicle exerts on the walls decreases as the level of
heterogeneity increases.}  As the compression becomes larger, and the shape more
pancake-like, both the high-curvature domains and base phase deform
significantly.  {A significant decrease occurs in the resisting force for the
heterogeneous vesicles as the concentration of the high-curvature species
increases, see Figure~\ref{fig:max_force_Top_verses_nhc}.}  These results show some of
the ways in which the phase-separated domains can contribute to mechanical
responses of heterogeneous vesicles.

\subsection{Insertion and Transport of Heterogeneous Vesicles within Channels}
\label{sec:transport_experiment}

{The mechanics and kinetics of heterogeneous structures inserting into
small capillaries, pores, or channels play an important in cellular processes
and microfluidic devices~\cite{Li2017,Bertrand2012,Patty2003}.  Toward
understanding general principles, model physical systems have been 
studied~\cite{PatrickShelby2003,Benet2016}.
In~\cite{PatrickShelby2003}, a microfluidics model system was developed 
to investigate the role of elastic mechanics of red blood cells in traversing
micro-capillaries when in healthy states and when in diseased states, such as
malaria which has increased rigidity.  In~\cite{Benet2016}, the permeation of
homogeneous vesicles through narrow pores was studied with an emphasis on the
roles of surface adhesion, elasticity, and surface tension on the pressure
differences required to drive transport through pores.  }

\begin{figure}[H] \centering
\includegraphics[width=0.99\columnwidth]{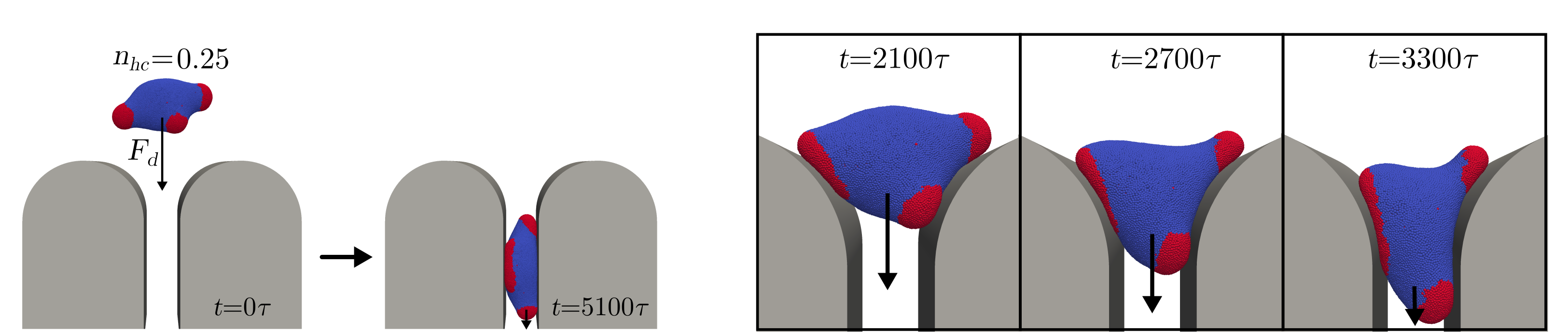}
\caption{{Heterogeneous vesicles inserting into channels.
Vesicles experience a constant pressure load force
$\boldsymbol{F}_d=-(0.015\epsilon/\sigma)\hat{\boldsymbol{z}}$
that drives insertion and transport within slit-like channels.
The channels have wall-separation distance
$30\sigma$ with an entrance smoothed by a cylinder having radius $50\sigma$  
\textit{(left)}.  The vesicle significantly deforms and rearranges phase domains 
during insertion into the channel opening \textit{(right)}.  
Results for vesicle transit times are shown in Figure
\ref{fig:passage_delay}.}}
\label{fig:passage_diagram} 
\end{figure}

{We investigate the behaviors of heterogeneous vesicles 
when inserted and transported within slit-like channels.
The vesicle kinetics of insertion and transport are studied 
when varying the concentration of the high-curvature phase.
The hydrodynamics is treated as a simplified model with a 
constant drag and pressure force acting 
on all particles of the vesicle.  This is motivated by 
the hypothesis that as opposed to flow the insertion 
process is driven primarily by surface pressures of 
the incompressible fluid transmitted over the membrane 
surface of the vesicles.  Other approaches, such as 
fluctuating hydrodynamics methods incorporating more 
detailed fluid mechanics could also be used to extend 
our coarse-grained model as 
in~\cite{AtzbergerLAMMPS2016}.  The 
driving force is taken to be $\boldsymbol{F}_d =
-(0.015\epsilon/\sigma)\hat{\boldsymbol{z}}$.  The channel geometry consists of
two plates with separation distance $30\sigma$.  The channel entrance is
smoothed by two cylinders, both of radius $50\sigma$.}  At the start, vesicles
have effective radii ranging from $r_0 \sim 30.1\sigma$ for $n_{hc} = 0.0$ to
$r_0 \sim 26.0\sigma$ for $n_{hc} = 0.3$, as determined from the $a_{00}$
basis-expansion coefficient corresponding to spherical harmonic mode $Y_0^0$.
{The channel and some typical configurations of the vesicles before,
during, and after insertion are shown in Figure~\ref{fig:passage_diagram}.}

\begin{figure}[H] \centering
\includegraphics[width=0.5\columnwidth]{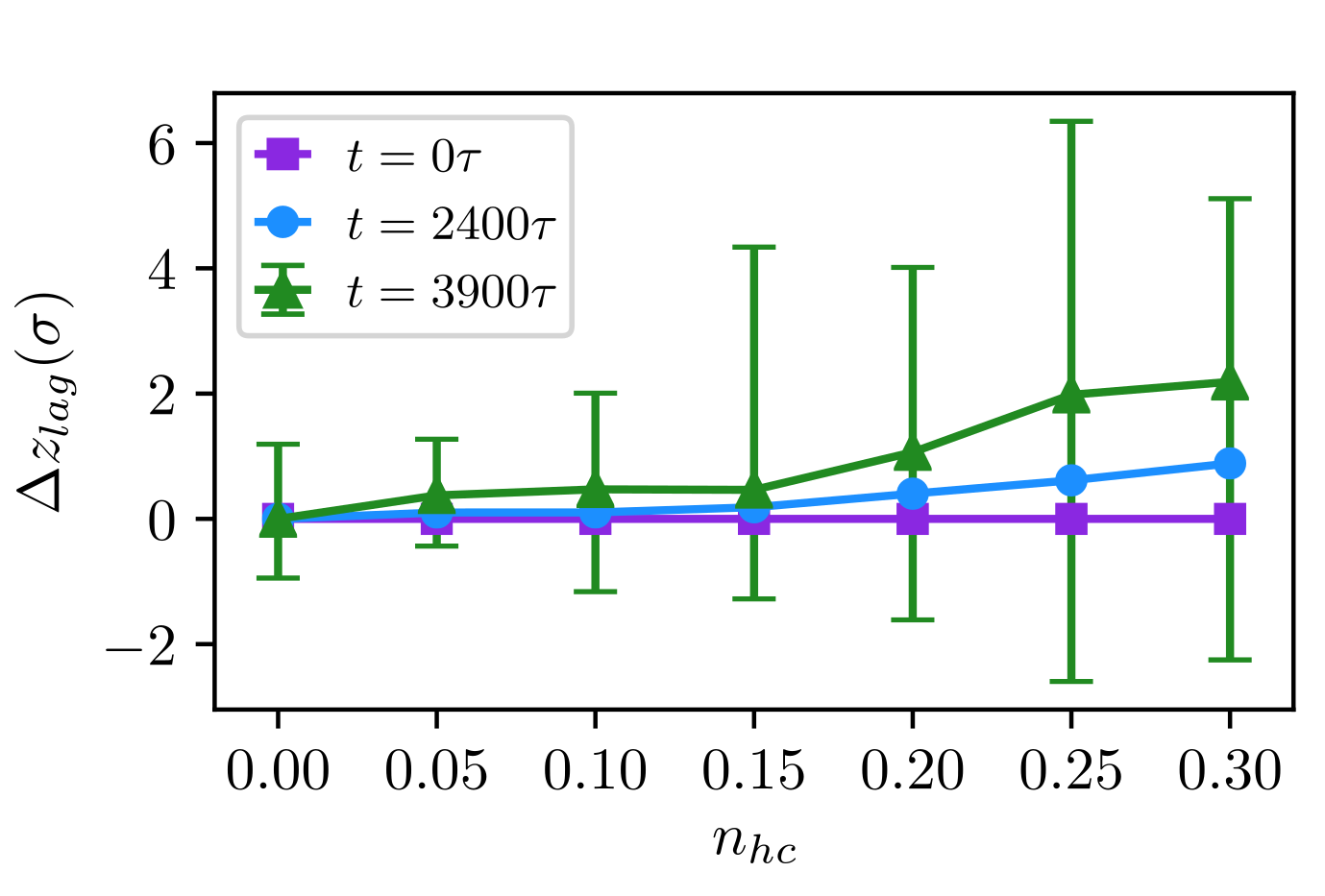}
\caption{{The vesicle shape changes during insertion into channels.
For vesicles with different phase mixtures $n_{hc}$, the distance $z(t)$ 
is shown for how far a vesicle has traveled down the channel
by time $t$.  The lag-distance
$\Delta{z}_{lag} = z(t;n_{hc}) - z_0(t)$ is reported for  
the position $z(t;n_{hc})$ of
the heterogeneous vesicle compared with the homogeneous case $z_0(t)$.  Results 
are averaged over $10$ simulations for vesicles with random initial
orientations.  For the case $t=3900\tau$, when all vesicles were fully inserted
into the channel, bars are shown indicating the range of the smallest and
largest lags observed over the trials.}} 
\label{fig:passage_delay} 
\end{figure}

{We compare for homogeneous and heterogeneous vesicles how the different
concentrations of the high-curvature phase impact the distance $z(t)$ the
vesicles are transported within the channel over time duration $t$.  The
distance of the homogeneous vesicle is denoted by $z_0(t)$.  We report the 
lag-distance $\Delta{z}_{lag} = z(t;n_{hc}) - z_0(t)$ between the position
$z(t;n_{hc})$ of the heterogeneous vesicle and the homogeneous case.  The
trials are averaged over $10$ random initial orientations for each vesicle.  We
report our results in Figure~\ref{fig:passage_delay}.}

\begin{figure}[H] \centering
\includegraphics[width=0.9\columnwidth]{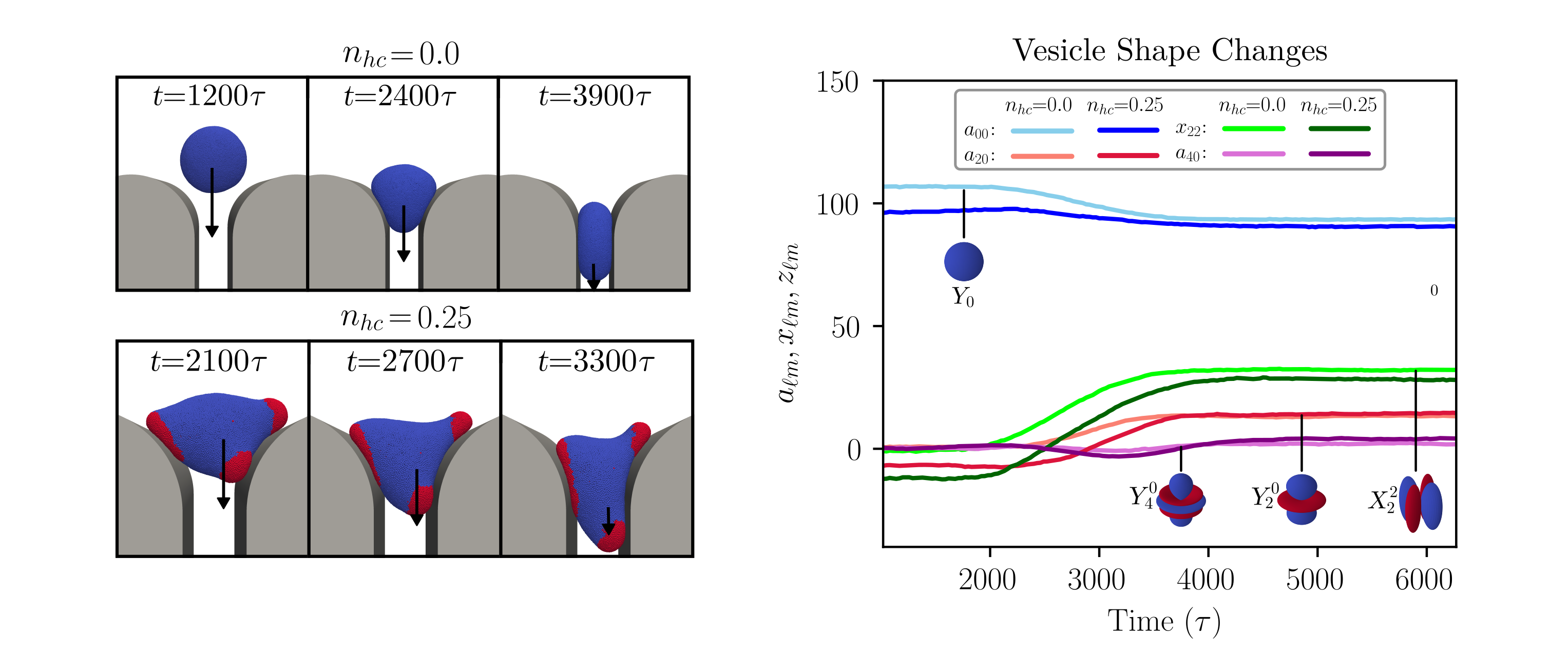}
\caption{{The vesicle shape changes during insertion into channels.
The changes are compared for a 
heterogeneous vesicle relative to the homogeneous case \textit{(left)}.
The changes in shape of vesicles is characterized using 
the spherical harmonics with coefficients $a_{00}$, $a_{20}$, $x_{22}$, and
$a_{40}$.  The harmonics are shown as radial functions with blue positive and red negative
for the modes $Y_0^0$, $Y_2^0$, $X_2^2$, and $Y_4^0$ \textit{(right)} }}
\label{fig:passage_shape_transformation} \end{figure}

{It was found the heterogeneous vesicles had longer transport 
times and larger variances in the channel insertion studies.
These vesicles enter the channel through a combination of 
rearrangement of the phase domains, further deformation-driven 
phase coarsening, and changes in the shape.  Since the vesicles have 
non-spherical shapes, the initial orientations when encountering 
the channel entrance also can play a role, see Figure~\ref{fig:passage_diagram}. 
From our discrete-to-continuum mappings these shape changes 
were quantified in terms of modes in Figure~\ref{fig:passage_shape_transformation}.
Relative to the homogeneous case, we found there were significant differences 
seen in the shapes during the insertion phase.  By the time $t = 3900\tau$, 
all of the vesicles were fully inserted within the channel and took on 
similar overall ellipsoidal shapes.  For heterogeneous vesicles, this 
combination of effects were found to lead to significant differences 
in transport times and insertion kinetics relative to the homogeneous 
case, see Figs.~\ref{fig:passage_diagram}--\ref{fig:passage_delay}.
}

{It was interesting that while the heterogeneous vesicles 
could better accommodate deformations during compression incurring 
a lower energetic cost, they still took more time on average to insert 
into channels.  This appears to be a
consequence of the sequential nature of the insertion process.  During insertion
only the leading part of the vesicle is initially subjected to deformation.  
Depending on the initial vesicle orientation and arrangement of the phase domains, 
this could delay insertion, given the need for phase domains to rearrange 
to accommodate the deformations required for full insertion into the channel.
This is seen in the large variability of the insertion times, with the 
heterogeneous case sometimes occurring more rapidly or much more slowly 
than the homogeneous case, see Figure~\ref{fig:passage_delay}.  The
simulation results for the heterogeneous vesicles show how many of the 
behaviors differ significantly relative to their homogeneous 
counter-parts.}

\section{Conclusions}
{Heterogeneous vesicles with membranes having phase-separated domains
can exhibit interesting mechanical responses and other 
behaviors differing significantly from homogeneous vesicles.
Our coarse-grained simulation and analysis methods allow for 
better understanding the roles played by the phase-separated domains
both in passive shape fluctuations and during active deformations.
Our discrete-to-continuum mapping methods and spherical harmonics
approaches allow for relating 
configurations in coarse-grained descriptions to corresponding
continuum fields providing quantitative characterizations. 
For shape fluctuations
when varying the vesicle sizes, we showed how this could be 
used to make comparisons with theory based on continuum 
mechanics.  We further showed the phase-separated
domains break rotational symmetries resulting in passive shape 
fluctuations and two-point correlations having enhanced amplitude 
and variability relative to the homogeneous case.  We also showed how our 
approaches can be used to probe mechanical responses 
when subjected to more active deformations.
When compressing vesicles between two plates, we found the 
high-curvature phases can rearrange to accommodate bending stresses
by distributing near the most curved edges of the compressing vesicle.  For studies
of vesicle insertion into channels, we further found that the heterogeneity
can have mixed results.  When the orientation of the vesicle had 
the ellipsoidal major axis aligned with the channel, we found 
heterogeneous vesicles can insert rapidly.  However, 
when the orientation is orthogonal, we found delays can 
arise from deformations and rearrangements of the phase-separated 
domains during vesicle insertion into the channel.  This manifested as significant 
variability in transport times.  These results suggest a few novel mechanisms 
by which membrane heterogeneity can 
augment membrane mechanics and kinetics.  Our introduced methods provide
ways to quantitatively characterize these differences.  Our methods 
provide general approaches 
for further investigations of phenomena
within heterogeneous membranes taking into account the roles played by 
phase separation, thermal fluctuations, geometry, and mechanics.
Our results show that the presence of even a modest number of 
phase-separated domains in heterogeneous membranes
can significantly augment their mechanical responses relative to the 
homogeneous case.
}

\section*{Acknowledgments}

The authors D.A.R., and P.J.A acknowledge support from research grants NSF Grant
DMS - 1616353, DOE ASCR CM4 DE-SC0009254, and DOE Grant ASCR PHILMS
DE-SC0019246.  P.J.A. also would like to acknowledge a hardware grant from
Nvidia.  D.A.R. also would like to thank B. Gross for a few helpful technical
discussions.  We also acknowledge UCSB Center for Scientific Computing NSF DMR
1720256 and NSF CNS 1725797.

\printbibliography

\appendix

\section{Gradients of the Energy $U(\mb{r}_{ij},\mb{n}_i,\mb{n}_j)$}
\label{appendix:gradients}
{The detailed expressions are given for each of the gradients of the potential
energy with respect to the translational $\mathbf{r}$ and rotational
$\mathbf{n}_i$ degrees of freedom contributing in equation\ref{Potential}.}  These
gradients can be expressed as
\begin{eqnarray}
\nabla_{\mathbf{r}_i}U &=&
\begin{cases}
  \nabla_{\mathbf{r}_i}u_{R}(r)-\epsilon\mu\nabla_{\mathbf{r}_i}a & \quad
    r<r_{b} \\
  \phi\nabla_{\mathbf{r}_i}u_{A}(r)+u_{A}(r)\mu\nabla_{\mathbf{r}_i}a & \quad
    r_{b}<r<r_{c}
\end{cases} \\
\nabla_{\mathbf{n}_i}U &=&
\begin{cases}
    -\epsilon\mu\nabla_{\mathbf{n}_i}a & \quad r<r_{b} \\
    u_{A}(r)\mu\nabla_{\mathbf{n}_i}a & \quad r_{b}<r<r_{c}
\end{cases}\\
\nabla_{\mathbf{r}_i}u_{R}(r) &=&
-\frac{4\epsilon}{r_{b}}\bigg[\Big(\frac{r_{b}}{r}\Big)^{5}-
  \Big(\frac{r_{b}}{r}\Big)^3\bigg] \mathbf{\hat{r}}_{ij}
\label{grad_r_u_R}\\
\nabla_{\mathbf{r}_i}u_{A}(r) &=&
\frac{\epsilon\pi\zeta}{(r_{c}-r_{b})}\cos^{2\zeta-1}(\xi)\sin(\xi)
  \mathbf{\hat{r}}_{ij},
  \quad \xi = \frac{\pi}{2}\frac{(r-r_{b})}{(r_{c}-r_{b})}
\label{grad_r_u_A} \\
\nabla_{\mathbf{r}_i}a & = &
\frac{1}{r}\big[2(\mathbf{n}_{i}\cdot\mathbf{\hat{r}}_{ij})
  (\mathbf{n}_{j}\cdot\mathbf{\hat{r}}_{ij})\mathbf{\hat{r}}_{ij}-
  (\mathbf{n}_{i}\cdot\mathbf{\hat{r}}_{ij})\mathbf{n}_{j}-
  (\mathbf{n}_{j}\cdot\mathbf{\hat{r}}_{ij})\mathbf{n}_{i}  \\
\nonumber
&- & \sin\theta_{0}\big(\mathbf{n}_{j}-\mathbf{n}_{i}-
  [(\mathbf{n}_{j}-\mathbf{n}_{i})\cdot\mathbf{\hat{r}}_{ij}]
  \mathbf{\hat{r}}_{ij}\big)\big]
\label{grad_r_phi}\\
  \nabla_{\mathbf{n}_i}a &=&
    \mathbf{n}_{j}-(\mathbf{n}_{j}\cdot\mathbf{\hat{r}}_{ij}-
    \sin\theta_{0})\mathbf{\hat{r}}_{ij}.
\label{grad_ni_phi}
\end{eqnarray}

\section{Computing the Wall-Force: Resistance of Vesicles to Compression}
\label{appendix:compression}
For compression of the vesicles between the two walls, we compute the effective
wall-force, related to the pressure, exerted by the vesicle on the walls.  
{Consider the geometry of a wall $w$ spanning the $xy$-plane at $z = z_{w}$.  Let
$z_i$ denote the $z$-position of the $i^{th}$ particle.  Consider the
free-energy of the particle-wall interaction $U_w(z_i-z_w)$, which in our
studies is a 9-3 LJ-potential.}  The force exerted by the $i^{th}$ particle on
the wall is given by $\tilde{F}_{w, i} = -\nabla_{z_w} U_w(z_i-z_w)$. For our
studies of compression, we sum over all of the vesicle's particles with $|z_i -
z_w| < 2.5\sigma$ to obtain the total particle-wall force
\begin{equation}
F_{w} = -\sum_{i} \nabla_{z_w} U_w(z_i-z_w).
\end{equation}
In our studies, the bottom wall is stationary, so we use the wall separation
$z = z_{Top} - z_{Bottom}$ to parameterize the force $F_{Top}(z)$. We use this
approach to compute the reported resisting forces when compressing homogeneous
and heterogeneous vesicles between the planar walls in
Section~\ref{sec:compression_experiment}.

\end{document}